\documentclass[]{llncs}

\usepackage{pslatex}
\usepackage{latexsym}
\usepackage{amsxtra} 
\usepackage{amssymb}
\usepackage{amsmath}
\usepackage{stmaryrd}
\usepackage{mathtools}
\usepackage{tikz}
\usetikzlibrary{arrows,decorations.pathmorphing,backgrounds,positioning,fit,petri,calc,automata}
\usetikzlibrary{calc,decorations.pathreplacing}

\usepackage{algorithm}
\usepackage{algorithmicx}
\usepackage[noend]{algpseudocode}

\usepackage{soul}
\usepackage{color}
\usepackage{times}

\usepackage{graphics}
\usepackage{epsfig}
\usepackage{wrapfig}

\usepackage{multirow}

\usepackage{setspace}
\usepackage[noding,draft]{commenting}

\renewcommand{\iff}{\Leftrightarrow}

\DeclareMathOperator{\llcm}{lcm}

\DeclareMathOperator{\positionsOp}{positions}
\DeclareMathOperator{\verticesOp}{vertices}

\DeclareMathOperator{\segmentsOp}{segments}
\DeclareMathOperator{\lbsegmentsOp}{lb-segments}
\DeclareMathOperator{\extentOp}{extent}

\DeclareMathOperator{\lcornersOp}{l-corners}
\DeclareMathOperator{\bcornersOp}{b-corners}
\DeclareMathOperator{\lbcornersOp}{lb-corners}
\DeclareMathOperator{\varsOp}{vars}
\DeclareMathOperator{\atomsOp}{atoms}

\DeclareMathOperator{\minpathOp}{min-path}
\DeclareMathOperator{\minweightOp}{min-weight}

\newcommand{\zedNval}[0]{\zed^{\x}}

\newcommand{\zedNNval}[0]{\zedNval \!\times\! \zedNval}

\newcommand{\x}[0]{\vec{x}}
\newcommand{\xk}[1]{\vec{x}^{(#1)}}

\newcommand{\xxki}[2]{x^{(#1)}_{#2}}
\newcommand{\y}[0]{\vec{y}}

\newcommand{\z}[0]{\vec{z}}

\newcommand{\mi}[0]{-\!}

\newcommand{\biinf}[1]{{}^\infty\!{#1}^\infty}

\newcommand{\bin}[1]{{\llen{#1}}_2}

\newcommand{\abs}[1]{\mbox{abs}(#1)}

\newcommand{\rbr}{{\bf ]\!]}}
\newcommand{\lbr}{{\bf [\![}}
\newcommand{\sem}[1]{\lbr #1 \rbr}

\renewcommand{\vec}[1]{{\bf {#1}}}

\newcommand{\true}{\mbox{\bf true}}
\newcommand{\false}{\mbox{\bf false}}
\newcommand{\atoms}[1]{\atomsOp(#1)}

\newcommand{\len}[1]{{|{#1}|}}
\newcommand{\llen}[1]{{|\!|{#1}|\!|}}

\newcommand{\card}[1]{\mbox{card}({#1})}

\newcommand{\weight}[1]{\omega(#1)}
\newcommand{\vertices}[1]{\verticesOp(#1)}
\newcommand{\positions}[1]{\positionsOp(#1)}

\newcommand{\segments}[1]{\segmentsOp(#1)}
\newcommand{\lbsegments}[1]{\lbsegmentsOp(#1)}
\newcommand{\extent}[1]{\extentOp(#1)}

\newcommand{\lcorners}[1]{\lcornersOp(#1)}
\newcommand{\bcorners}[1]{\bcornersOp(#1)}
\newcommand{\lbcorners}[1]{\lbcornersOp(#1)}
\newcommand{\vars}[1]{\varsOp(#1)}

\newcommand{\minweight}[1]{\minweightOp({#1})}

\newcommand{\arrow}[2]{\xrightarrow[{\scriptstyle #2}]{{\scriptstyle #1}}}

\newcommand{\nat}{{\bf \mathbb{N}}}
\newcommand{\zed}{{\bf \mathbb{Z}}}

\def\vr{\kern-\arraycolsep & \kern-\arraycolsep}
\def\VR{\kern-\arraycolsep\strut\vrule}

\newif\ifLongVersion\LongVersiontrue
\usepackage{arrayjobx}
\newcommand{\gridName}[2]{n#1n#2}
\newcommand{\gridGen}[9]
{

  \tikzset{
    nborder/.style={circle,inner sep=0.5mm,minimum size=0mm},
    ngrid/.style={nborder,fill=black!30,draw=black}
  }

  \begin{scope}

    \def\xIndent{#7}
    \def\yIndent{#8}

    \pgfmathsetmacro{\aa}{#1}
    \foreach \y in {1,...,#6} {
      \pgfmathtruncatemacro\inxy{#6-\y+1}
      \def\nName{\gridName{0}{\inxy}}
      \pgfmathsetmacro{\bb}{#4+(\y-1)*#5}
      \node (\nName) at (\aa cm,\bb cm) [nborder] {$x_{\inxy}$};
    }
    \pgfmathsetmacro{\bb}{#4+(#6-1)*#5+\yIndent}
    \foreach \x in {1,...,#3} {
      \pgfmathsetmacro{\aa}{\xIndent+#1+(\x-1)*#2}
      \def\nName{\gridName{\x}{0}}
      \pgfmathtruncatemacro\inx{#9+\x-1}
      \node (\nName) at (\aa cm,\bb cm) [nborder] {$\vec{x}^{(\inx)}$};
    }
    \foreach \x in {1,...,#3} {
      \pgfmathsetmacro{\aa}{\xIndent+#1+(\x-1)*#2}
      \foreach \y in {1,...,#6} {
        \pgfmathtruncatemacro\inxy{#6-\y+1}
        \pgfmathsetmacro{\bb}{#4+(\y-1)*#5}
        \def\nName{\gridName{\x}{\inxy}}
        \node (\nName) at (\aa cm,\bb cm) [ngrid] {};
      }
    }

  \end{scope}
}

\newcommand{\gridEdge}[4]
{
  \draw [->,thick] (\gridName{#1}{#2}) -- (\gridName{#3}{#4});
}

\newcommand{\gridEdgeLStyle}[6]
{
  \draw [->,thick] (\gridName{#1}{#2}) -- node[#6] {$#5$} (\gridName{#3}{#4});
}

\tikzset{
  nborder/.style={circle,inner sep=0.75mm,minimum size=0mm},
  ngrid/.style={nborder,fill=black!30,draw=black},
  matchingsep/.style={shape=circle,fill=red!10,inner sep=0mm,minimum size=0mm},
  matching/.style  = {matchingsep,midway,left}
}

\tikzset{
  zrBorder/.style={circle,inner sep=0.5mm,minimum size=0mm},
  zrNode/.style={zrBorder,fill=black!40} % draw=black
}

\tikzset{
  nborder/.style={circle,inner sep=0.5mm,minimum size=0mm},
  ngrid/.style={nborder,fill=black!40},
  matchingsep/.style={shape=circle,fill=red!10,inner sep=0mm,minimum size=0mm},
  matching/.style  = {matchingsep,midway,left}
}

\newcommand{\TermGridGenNoCap}[8]
{
  \begin{scope}
    \def\xIndent{#7}
    \def\yIndent{#8}
    \def\yBase{#4+(#6-1)*#5}
    \foreach \x in {1,...,#3} {
      \pgfmathsetmacro{\aa}{\xIndent+#1+(\x-1)*#2}
      \foreach \y in {1,...,#6} {
        \pgfmathsetmacro{\bb}{\yBase-(\y-1)*#5}
        \def\nName{\gridName{\x}{\y}}
        \node (\nName) at (\aa cm,\bb cm) [ngrid] {};
      }
    }
  \end{scope}
}

\newcommand{\TermGridGen}[9]
{

  \begin{scope}

    \def\xIndent{#7}
    \def\yIndent{#8}

    \pgfmathsetmacro{\aa}{#1}
    \foreach \y in {1,...,#6} {
      \pgfmathsetmacro{\bb}{#4+(\y-1)*#5}
      \def\nName{\gridName{0}{\y}}
      \pgfmathtruncatemacro\inx{#6-\y+1}
      \node (\nName) at (\aa cm,\bb cm) [nborder] {$x_{\inx}$};
    }
    \pgfmathsetmacro{\bb}{#4+(#6-1)*#5+\yIndent}
    \foreach \x in {1,...,#3} {
      \pgfmathsetmacro{\aa}{\xIndent+#1+(\x-1)*#2}
      \def\nName{\gridName{\x}{0}}

      \pgfmathtruncatemacro\inxvv{mod(\x-1,#3-1)}
      \pgfmathtruncatemacro\inxww{\x-1}
      \def\inx{\ifthenelse{\equal{#9}{1}}{\inxvv}{\inxww}}

      \node (\nName) at (\aa cm,\bb cm) [nborder] {$\vec{x}^{(\inx)}$};
    }

    \TermGridGenNoCap{#1}{#2}{#3}{#4}{#5}{#6}{#7}{#8}

  \end{scope}
}
\newcommand{\TermGridGenDifferentCap}[9]
{

  \begin{scope}

    \def\xIndent{#7}
    \def\yIndent{#8}

    \pgfmathsetmacro{\aa}{#1}
    \pgfmathsetmacro{\aap}{#1+(#2)*(#3-1)+2*#7}
    \foreach \y in {1,...,#6} {
      \pgfmathsetmacro{\bb}{#4+(\y-1)*#5}
      \def\nName{\gridName{0}{\y}}
      \pgfmathtruncatemacro\inx{#6-\y+1}
      \node (\nName) at (\aa cm,\bb cm) [nborder] {$x_{\inx}$};
      \node (\nName) at (\aap cm,\bb cm) [nborder] {$x'_{\inx}$};
    }

    \TermGridGenNoCap{#1}{#2}{#3}{#4}{#5}{#6}{#7}{#8}

  \end{scope}
}

\newcommand{\TermGridEdgeC}[6]
{
  \draw [->,thick] (\gridName{#1}{#2}) -- (\gridName{#3}{#4})
     node[midway,#6] {#5}; 
}

\tikzset{
  zrBorder/.style={circle,inner sep=0.5mm,minimum size=0mm},
  zrNode/.style={zrBorder,fill=black!40} % draw=black
}

\declareauthor{fk}{Filip}{red}
\declareauthor{vk}{Viktor}{green}
\declareauthor{ri}{Radu}{blue}

\title{PTIME Computation of Transitive Closures of Octagonal Relations}

\author{Filip Kone\v{c}n\'{y}} 
\institute{Swiss Federal Institute of Technology Lausanne (EPFL)}

\begin{document}

\maketitle

\begin{abstract}
Computing transitive closures of integer relations is the key to finding precise invariants of integer programs. In this paper, we study {\em difference bounds} and {\em octagonal} relations and prove that their transitive closure is a PTIME-computable formula in the existential fragment of Presburger arithmetic. This result marks a significant complexity improvement, as the known algorithms have EXPTIME worst case complexity.
\end{abstract}

%%%%%%%%%%%%%%%%%%%%%%%%%%%%%%%%%%%%%%%%%%%%%%%%%%%%%%%%%%%
%%%%%%%%%%%%%%%%%%%%%%%%%%%%%%%%%%%%%%%%%%%%%%%%%%%%%%%%%%%
\section{Introduction}

This paper gives the first polynomial-time algorithm for computing {\em closed forms} of {\em difference bounds} and {\em octagonal} relations. Difference bounds (DB) relations are relations defined as conjunctions over atomic propositions of the form $x-y\leq c$ where $c$ is an integer and $x,y$ range over unprimed and primed variables $\x\cup\x'$. Octagonal relations generalize difference bounds relation by allowing conjuncts of the form $\pm x \pm y \leq c$. Both classes of relations are widely used as domains in verification of software and hardware \cite{mine06,WangIGG05}.

Given a binary relation $R$ on states (represented as
a formula with primed and unprimed variables)
a \emph{closed form} $R$ is another formula $\widehat{R}(k)$
containing primed and unprimed variables as well as a
parameter variable $k$, such that substituting the parameter
$k$ with any integer $n \ge 1$ gives a precise description
of $R^n$, the $n$-th power of $R$.  The main result of this
paper is a polynomial-time algorithm that, given the formula $R$ in
the form of octagonal constraints computes a closed
form $\widehat{R}(k)$ as a formula in the
existential fragment of Presburger arithmetic. This result
immediately extends to the computation of an expression for transitive
closure, because $R^+ \iff \exists k\geq 1
~.~ \widehat{R}(k)$.

Approaches for computing the precise closed form of iterated relation compositions are referred to as
\emph{acceleration} algorithms.
Known acceleration algorithms for the two classes of relations are based on the notion of {\em periodicity} and compute closed forms of the size that is polynomial in the size of the {\em prefix} and the {\em period} of a relation, both of which can be exponentially large in the binary size of a relation $\bin{R}$.
Intuitively, $n$-th power of a DB relation $R$ can be obtained by computing minimal weights of paths between pairs of vertices in certain graphs (called {\em unfolded constraint graphs of $R$} and denoted $\mathcal{G}_R^n$). For a fixed pair, minimal weights evolve periodically as a function of $n$.
Due to these exponential bounds, 
an algorithm for computing closed forms that runs in time that is polynomial in $\bin{R}$ must necessarily be based on a method different than
explicitly computing periodicity. This paper presents the 
first such algorithm.

\subsubsection{Overview}

First, we study difference bounds relations (Section \ref{sec:dbrel} gives a background). Our main observation is that the problem of computing a closed form of a DB relation $R$ can be reduced to the computation of closed forms of two PTIME-computable DB relations $R_{fw}$ and $R_{bw}$ such that $R_{fw}$ ($R_{bw}$) belongs to a fragment called {\em forward} ({\em backward}) {\em one-directional} DB relations which contains DB relations of the form  
\[ \bigwedge_{ij} x_i-x_j'\leq c_{ij} \hspace{4mm} \big(\bigwedge_{ij} x_i'-x_j\leq c_{ij} \textrm{, respectively} \big) \]

We first study these (dual) fragments and give a PTIME algorithm which computes the closed form in the existential fragment of Presburger arithmetic (Section \ref{sec:fw:relations}). The main insight of this algorithm is that the closed form can be defined by encoding polynomially many {\em path schemes} which can be thought of as regular patterns that capture all paths with minimal weight in unfolded constraint graphs. 

Next, we observe that for a fixed pair of vertices $(u,v)$ in an unfolded constraint graph, any path $\rho$ from $u$ to $v$ can be {\em normalized}, i.e. replaced with another path $\rho'$ from $u$ to $v$ such that the weight of $\rho'$ is not greater than the weight of $\rho$ and $\rho'$ is in a normal form (Section \ref{sec:normalization}).
%$\weight{\rho'}\leq\weight{\rho}$ ($\weightOp$ denotes the weight of a path) 

Then, we define the relations $R_{fw}$ and $R_{bw}$ and show that there exists an integer $B$ of polynomial size such that every normalized path $\rho$ in $\mathcal{G}_R^{2B+n}$ can be viewed as a concatenation of several paths from $\mathcal{G}_{R_{fw}}^{n}$, $\mathcal{G}_{R_{bw}}^{n}$ and $\mathcal{G}_R^{B}$ (Section \ref{sec:closed:form:db}). Since paths from $\mathcal{G}_{R_{fw}}^{n}$ and $\mathcal{G}_{R_{bw}}^{n}$ are captured by closed forms $\widehat{R}_{fw}(n)$ and $\widehat{R}_{bw}(n)$ (both PTIME-computable), and since paths in $\mathcal{G}_R^{B}$ are captured by $R^B$ (also PTIME-computable, since $B$ is polynomially large), it follows that $\widehat{R}_{fw}(n)$, $\widehat{R}_{bw}(n)$, and $R^B$ can be combined to form a closed form $\widehat{R}(2B+n)$.

Finally, in Section \ref{sec:octagons}, we show that these methods and results can be generalized to compute closed forms of octagonal relations in polynomial time as well. Section \ref{sec:conclusions} concludes.

\subsubsection{Related work}

Octagonal constraints \cite{mine06} are well known in abstract interpretation as an abstract domain for over-approximating sets of reachable states. Transitive closure algorithms for octagonal relations \cite{cav10} are the core of reachability analysis techniques based on computation of procedure summaries \cite{fm12} or on accelerated interpolation \cite{atva12}.

DB and octagonal relations have been shown to have Presburger definable transitive closures \cite{comon-jurski,fundamenta,tacas09} and to have periodic  characterization \cite{cav10}. An algorithm from \cite{cav10} computes a transitive closure whose size is polynomial in the binary size of the relation $\bin{R}$ and in the size of the prefix and period. Since relations whose prefix or period increases exponentially in $\bin{R}$ can be constructed, the exponential lower bound on the size of the computed transitive closure follows.

Recently, \cite{vmcai14} proves that both prefix and period can also be upper-bounded by a single exponential and moreover, shows NP-completeness of the reachability problem for {\em flat counter systems}, a class of integer programs without nested loops where each loop (non-loop) transition is described by an octagonal relation (QFPA\footnote{Quantifier-Free Presburger Arithmetic} formula). Moreover, \cite{vmcai14} presents a non-deterministic reduction to satisfiability of QFPA formulas (an NP-complete problem), essentially by first guessing the prefix and period and then guessing one of exponentially many disjuncts of the transitive closure, for each loop. Our present result can turn this reduction into a deterministic one, since we can directly compute the transitive closure of each loop.

%Note that in the logic LIA* \cite{viktor-ruzica08}, which extends Presburger arithmetic, the star operator corresponds to a closure under vector addition. This operator cannot directly express transitive closure of difference bound relations, because not all difference bound relations correspond to adding vectors from a given Presburger arithmetic-definable set.

%%%%%%%%%%%%%%%%%%%%%%%%%%%%%%%%%%%%%%%%%%%%%%%%%%%%%%%%%%%
%%%%%%%%%%%%%%%%%%%%%%%%%%%%%%%%%%%%%%%%%%%%%%%%%%%%%%%%%%%

%%%%%%%%%%%%%%%%%%%%%%%%%%%%%%%%%%%%%%%%%%%%%%%%%%%%%%%%%%%
%%%%%%%%%%%%%%%%%%%%%%%%%%%%%%%%%%%%%%%%%%%%%%%%%%%%%%%%%%%
\section{Preliminary Definitions}
In the rest of this paper, let $N\geq 1$ and let $\x=\{x_1, x_2, ..., x_N\}$ be a~set of variables ranging over $\zed$. For each $n\in\zed$, we define a fresh copy of variables $\xk{n} \stackrel{def}{=} \{\xxki{n}{1},\dots,\xxki{n}{N}\}$. Similarly, $\x'$ denotes a fresh copy of primed variables $\x'=\{x_1',\dots,x_N'\}$.
We assume that the reader is familiar with Presburger arithmetic (PA).
%, and we denote by QFPA (quantifier-free Presburger arithmetic) the set of boolean combinations of linear inequalities and linear modulo constraints.
%
For a PA formula $\phi$, let $\atoms{\phi}$ denote the set of atomic propositions in $\phi$, and $\phi[t/x]$ denote the formula obtained by substituting the variable $x$ with the term $t$.
$\card{S}$ denotes the cardinality of a set $S$ and $\abs{c}$ denotes the absolute value of $c\in\zed$.
A \emph{valuation} of $\x$ is a function $\smash{\nu : \x \arrow{}{} \zed}$. The set of all such valuations is denoted by $\zed^{\x}$.
%, and we denote by $\zed^N$ the $N$-times cartesian product $\zed \times \ldots \times \zed$. 
%
Given a relation $R\subseteq \zed^{\x} \times \zed^{\x}$, we denote by $R^i$, for $i > 0$, the $i$-times composition of $R$ with itself. We denote by $R^+ = \bigcup_{i=1}^\infty R^i$ the {\em transitive closure} of $R$.
If $R(\x,\x')$ defines $R$, we denote by $R^n(\x,\x')$ a formula that defines the $n$-th power $R^n$.
A {\em closed form} of $R$ is a formula $\widehat{R}(k,\vec{x},\vec{x'})$, where $k \not\in \vec{x}$, such that $\widehat{R}[n/k]$ defines $R^n$, for all $n \geq 1$.
For a weighted graph $G$ and a pair of vertices $u,v$, we denote by $\minweight{u,v,G}$ the minimal weight over all paths from $u$ to $v$ in $G$.% (if there exists one).

%%%%%%%%%%%%%%%%%%%%%%%%%%%%%%%%%%%%%%%%%%%%%%%%%%%%%%%%%%%
%%%%%%%%%%%%%%%%%%%%%%%%%%%%%%%%%%%%%%%%%%%%%%%%%%%%%%%%%%%

%%%%%%%%%%%%%%%%%%%%%%%%%%%%%%%%%%%%%%%%%%%%%%%%%%%%%%%%%%%
%%%%%%%%%%%%%%%%%%%%%%%%%%%%%%%%%%%%%%%%%%%%%%%%%%%%%%%%%%%
\section{Difference Bounds Relations}
\label{sec:dbrel}
\begin{definition}\label{dbc}
A formula $\phi(\vec{x})$ is a~\emph{difference bounds (DB) constraint} if it is a~finite conjunction of atomic propositions of the form $x_i-x_j \le \alpha_{ij},~ 1 \le i,j \le N$, where $\alpha_{ij}\in\zed$. A~relation $R\subseteq\zedNNval$ is a~{\em difference bounds relation} if it can be defined by a~difference bounds constraint $\phi_R(\x,\x')$.
\end{definition}

%
%\begin{figure}[b]
\begin{wrapfigure}{r}{7cm}
\vspace{-15mm}
\begin{center}
\begin{tabular}{c}
 \begin{tabular}{cc}
  \mbox{\begin{minipage}{2.5cm}
      \scalebox{0.85}{\begin{tikzpicture}
        \TermGridGenDifferentCap{0.0}{1.0}{2}{0.0}{0.75}{4}{0.7}{0.5}{0}
        \foreach \ii in {1,...,1} {
          \pgfmathtruncatemacro\jj{\ii+1}
          \TermGridEdgeC{\ii}{2}{\jj}{1}{\tiny$-\!1$}{right}
          \TermGridEdgeC{\ii}{3}{\jj}{2}{\tiny$0$}{left}
          \TermGridEdgeC{\ii}{1}{\jj}{3}{\tiny$0$}{right}
          \TermGridEdgeC{\jj}{4}{\ii}{4}{\tiny$0$}{below}
          \TermGridEdgeC{\jj}{3}{\ii}{4}{\tiny$0$}{right}
        }
      \end{tikzpicture}}
  \end{minipage}} & 
  \mbox{\begin{minipage}{3.5cm}
      \scalebox{0.85}{\begin{tikzpicture}
        \TermGridGen{0.0}{1.0}{4}{0.0}{0.75}{4}{0.7}{0.5}{0}
        \foreach \ii in {1,...,3} {
          \pgfmathtruncatemacro\jj{\ii+1}
          \TermGridEdgeC{\ii}{2}{\jj}{1}{\tiny$-\!1$}{right}
          \TermGridEdgeC{\ii}{3}{\jj}{2}{\tiny$0$}{left}
          \TermGridEdgeC{\ii}{1}{\jj}{3}{\tiny$0$}{right}
          \TermGridEdgeC{\jj}{4}{\ii}{4}{\tiny$0$}{below}
          \TermGridEdgeC{\jj}{3}{\ii}{4}{\tiny$0$}{right}
        }
      \end{tikzpicture}}
  \end{minipage}}
  %\vspace{2mm}
  \\
  (a) $\mathcal{G}_R$ &
  (b) $\mathcal{G}_R^3$
 \end{tabular}
\end{tabular}
\end{center}
\vspace{-3mm}
\caption{The constraint graph $\mathcal{G}_R$ and its 3-times unfolding $\mathcal{G}_R^3$ for a~difference bounds relation 
  $R \iff x_2\mi x'_1\leq -1 \wedge x_3\mi x'_2\leq 0 \wedge
  x_1\mi x'_3\leq 0 \wedge x'_4\mi x_4\leq 0 \wedge x'_3\mi x_4\leq
  0$. }
\label{fig:dbrel}
\vspace{-10mm}
\end{wrapfigure}
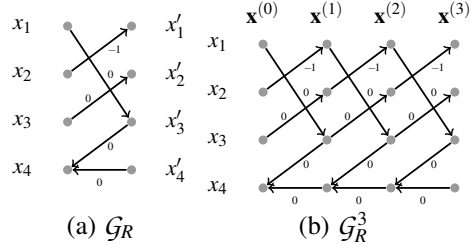

Difference bounds constraints are represented as graphs. If $\phi(\vec{x})$ is a~difference bounds constraint, then {\em constraint graph} $\mathcal{G}_\phi = \langle \vec{x}, \rightarrow \rangle$ is a weighted graph, where each vertex corresponds to a~variable, and there is an edge $x_i \arrow{\alpha_{ij}}{} x_j$ in $\mathcal{G}_\phi$ if and only if there exists a~constraint $x_i - x_j \leq \alpha_{ij}$ in $\phi$ (Fig.~\ref{fig:dbrel}(a)). The following result on existential quantification is well known \cite{mine-thesis}:
%It is well-known that DB constraints have existential quantification (using the classical Floyd-Warshall minimal weight path algorithm).
%
\begin{proposition} \label{dbm:exist:elim}
Let $\phi(x_0,\x)$ be a DB constraints. Then, $R(x_0,\x)$ is consistent if and only if $\mathcal{G}_\phi$ contains no negative cycle. If $\phi(x_0,\x)$ is consistent, then 
\[ \exists x_0 ~.~ \phi(x_0,\x) \iff \bigwedge_{x,y\in\x} x-y\leq \minweight{x,y,\mathcal{G}_\phi} \] 
%where $\min_{[\mathcal{G}_\phi]}\{x \arrow{}{} y\}$ stands for the minimal weight over all paths from $x$ to $y$ in $\mathcal{G}_\phi$.
Moreover, consistency check and computation of $\exists x_0 ~.~ \phi(x_0,\x)$ is in $\mathcal{O}(\bin{R})$ time.
\end{proposition}
Consequently, DB relations are closed under relational composition, i.e. $R^n(\x,\x')$ is a DB contraint for all $n\geq 1$. The $n$-th power of $R$ can be seen as a constraint graph consisting of $n$ copies of $\mathcal{G}_R$ (see Fig.~\ref{fig:dbrel}(b)):
\begin{definition}
Let $n\geq 1$ and $R(\x,\x')$ be a DB constraint. Then, the {\em $n$-times unfolding} of $\mathcal{G}_R$ is defined as $\mathcal{G}_R^n \stackrel{def}{=} \bigcup_{i=0}^{n-1} \mathcal{G}_{R(\xk{i},\xk{i+1})}$.
%$\mathcal{G}_R^n \stackrel{def}{=} \mathcal{G}_{T}$, where $T$ is a DB constraint $T \stackrel{def}{=} \bigwedge\limits_{i=1}^{n} R(\x,\x')[\xk{i-1}/\x,\xk{i}/\x']$.
\end{definition}
The vertices $\xk{0}\cup\xk{n}$ of $\mathcal{G}_R^n$ are called {\em extremal}. A path in $\mathcal{G}_R^n$ is said to be {\em extremal} if its first and last vertex are both extremal.
The next lemma gives means to compute $R^n(\x,\x')$ and test its consistency, by analyzing extremal paths of $\mathcal{G}_R^n$.
\begin{lemma} \label{computing:powers} (Lemma 6 in \cite{arxiv-vmcai14})
Let $n\geq 1$ and $R(\x,\x')$ be a DB constraints. Then, $R^n(\x,\x')$ is consistent if and only if $\mathcal{G}_R^n$ contains no extremal cycle with negative weight. If $R^n(\x,\x')$ is consistent, then $R^n(\x,\x')$ can be computed as
%\begin{equation}\label{dbm-min-paths}
\[
\begin{array}{rcl}
\bigwedge_{\scriptscriptstyle{1 \leq i, j \leq N}} & x_i - x_j \leq
\minweight{ \xxki{0}{i}, \xxki{0}{j}, \mathcal{G}_R^n } \wedge
x'_i - x'_j \leq \minweight{ \xxki{n}{i}, \xxki{n}{j}, \mathcal{G}_R^n } ~\wedge \\ 
& x_i - x_j' \leq \minweight{ \xxki{0}{i}, \xxki{n}{j}, \mathcal{G}_R^n } \wedge 
x'_i - x_j \leq \minweight{ \xxki{n}{i}, \xxki{0}{j}, \mathcal{G}_R^n }
\end{array}
\]
%\end{equation}
\noindent
%where $\min_{[\mathcal{G}_R^n]}\{\xxki{p}{i} \arrow{}{} \xxki{q}{j}\}$ stands for the minimal weight over all paths from $\xxki{p}{i}$ to $\xxki{q}{j}$ in $\mathcal{G}_R^n$, for $p,q \in \{0,n\}$. 
Moreover, consistency check and computation of $R^n(\x,\x')$ is in $\mathcal{O}(\bin{R}\cdot\log_2n)$ time.
\end{lemma}

%%%%%%%%%%%%%%%%%%%%%%%%%%%%%%%%%%%%%%%%%%%%%%%%%%%%%%%%%%%
\subsubsection{Paths in Unfoldings of $\mathcal{G}_R$}
%%%%%%%%%%%%%%%%%%%%%%%%%%%%%%%%%%%%%%%%%%%%%%%%%%%%%%%%%%%

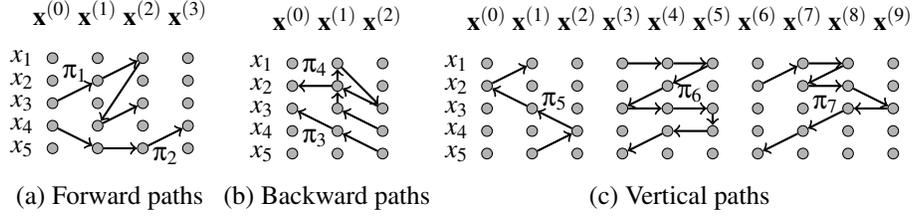
\begin{figure}[htbp]
\begin{center}
\begin{tabular}{ccc}
\begin{tikzpicture}[scale=1.0]
  % styles
  \tikzset{
    matchingsep/.style={shape=circle,fill=red!10,inner sep=0mm,minimum size=0mm},
    matching/.style  = {matchingsep,midway,left}
  }

  \gridGen{0.0}{0.6}{4}{0.0}{0.3}{5}{0.4}{0.6}{0}
  % path 1
  \gridEdgeLStyle{1}{3}{2}{2}{\pi_1}{above}
  \gridEdge{2}{2}{3}{1}
  \gridEdge{3}{1}{2}{4}
  \gridEdge{2}{4}{3}{3}
  % path 2
  \gridEdge{1}{4}{2}{5}
  \gridEdge{2}{5}{3}{5}
  \gridEdgeLStyle{3}{5}{4}{4}{\pi_2}{below}
\end{tikzpicture}
&
\begin{tikzpicture}[scale=1.0]
  % styles
  \tikzset{
    matchingsep/.style={shape=circle,fill=red!10,inner sep=0mm,minimum size=0mm},
    matching/.style  = {matchingsep,midway,left}
  }

  \gridGen{0.0}{0.6}{3}{0.0}{0.3}{5}{0.4}{0.6}{0}
  % path 3
  \gridEdge{3}{5}{2}{4}
  \gridEdgeLStyle{2}{4}{1}{3}{\pi_3}{below}
  % path 7
  \gridEdgeLStyle{2}{2}{1}{2}{\pi_4}{above}
  \gridEdge{2}{2}{2}{1}
  \gridEdge{2}{1}{3}{3}
  \gridEdge{3}{3}{2}{2}
  \gridEdge{2}{3}{2}{2}
  \gridEdge{3}{4}{2}{3}
\end{tikzpicture}
&
\begin{tikzpicture}[scale=1.0]
  % styles
  \tikzset{
    matchingsep/.style={shape=circle,fill=red!10,inner sep=0mm,minimum size=0mm},
    matching/.style  = {matchingsep,midway,left}
  }

  \gridGen{0.0}{0.6}{10}{0.0}{0.3}{5}{0.4}{0.6}{0}
  % path 4
  \gridEdge{2}{5}{3}{4}
  \gridEdgeLStyle{3}{4}{2}{3}{\pi_5}{above}
  \gridEdge{2}{3}{1}{2}
  \gridEdge{1}{2}{2}{1}
  % path 5
  \gridEdge{4}{1}{5}{1}
  \gridEdge{5}{1}{6}{1}
  \gridEdgeLStyle{6}{1}{5}{2}{\pi_6}{below}
  \gridEdge{5}{2}{4}{3}
  \gridEdge{4}{3}{5}{3}
  \gridEdge{5}{3}{6}{3}
  \gridEdge{6}{3}{6}{4}
  \gridEdge{6}{4}{5}{4}
  \gridEdge{5}{4}{4}{5}
  % path 6
  \gridEdge{7}{2}{8}{1}
  \gridEdge{8}{1}{9}{1}
  \gridEdge{9}{1}{8}{2}
  \gridEdgeLStyle{8}{2}{9}{2}{\pi_7}{below}
  \gridEdge{9}{2}{10}{3}
  \gridEdge{10}{3}{9}{3}
  \gridEdge{9}{3}{8}{4}
  \gridEdge{8}{4}{7}{5}
\end{tikzpicture}
\\
(a) Forward paths
&
(b) Backward paths
& 
(c) Vertical paths
\end{tabular}
\caption{Paths in unfolded constraint graphs. Path $\pi_1$ is repeating and essential, $\pi_2$ is repeating, $\pi_3$ is essential, $\pi_4$ is neither essential nor elementary, $\pi_5$ is essential. We have $\len{\pi_1}=4, \len{\pi_2}=3$, $\llen{\pi_2}=3, \llen{\pi_1}=\llen{\pi_3}=\llen{\pi_4}=2, \llen{\pi_5}=\llen{\pi_6}=\llen{\pi_7}=0$, $\positions{\pi_6}=\{3,4,5\}, \vars{\pi_2}=\{x_4,x_5\}, \vertices{\pi_2}=\{\xxki{0}{4},\xxki{1}{5},\xxki{2}{5},\xxki{3}{4}\}$.}
\label{figure:paths:in:unfolded}
\end{center}
\vspace{-5mm}
\end{figure}

In this paper, when the exact number of iterations does not matter, we sometimes consider paths in the bi-infinite unfolding $\biinf{\mathcal{G}_R}$ of $\mathcal{G}_R$, defined as
\[ \biinf{\mathcal{G}_R} \stackrel{def}{=} \bigcup_{i\in\zed} \mathcal{G}_{R(\xk{i},\xk{i+1})}\]
Note that each edge in $\biinf{\mathcal{G}_R}$ is either {\em forward} (i.e. of the form $\xxki{p}{i}\arrow{\alpha}{}\xxki{p+1}{j}$ for some $1\leq i,j\leq N$ and $p,\alpha\in\zed$), {\em backward} ($\xxki{p+1}{i}\arrow{\alpha}{}\xxki{p}{j}$), or {\em vertical} ($\xxki{p}{i}\arrow{\alpha}{}\xxki{p}{j}$).
A {\em path} is a sequence of the form (see Fig. \ref{figure:paths:in:unfolded} for illustrations) \[ \rho = \xxki{p_0}{i_0}\arrow{\alpha_0}{}\xxki{p_1}{i_1}\arrow{\alpha_1}{}\dots\arrow{\alpha_{n-1}}{}\xxki{p_n}{i_n} \] for some $n\geq 0$ where $\xxki{p_k}{i_k}\arrow{\alpha_k}{}\xxki{p_{k+1}}{i_{k+1}}$ is an edge in $\biinf{\mathcal{G}_R}$, for each $0\leq k<n$. 
We say that a variable $x_{i_k}$ {\em occurs} on $\rho$ at {\em position} $p_k$, for each $0\leq k\leq n$.
We say that $\rho$ is {\em forward} ({\em backward}, {\em vertical}) if $p_0<p_n$ ($p_0>p_n$, $p_0=p_n$, respectively).
The {\em length} and {\em relative length} of $\rho$ is defined as $\len{\rho}=n$ and $\llen{\rho}=\abs{p_n-p_0}$.
The {\em weight} of $\rho$ is defined as $\weight{\rho}=\alpha_0+\dots+\alpha_{n-1}$.
%The {\em weight} and {\em relative weight} of $\rho$ is defined as $\weight{\rho}=\alpha_0+\dots+\alpha_{n-1}$ and $\awwPar{\rho}=\frac{\weight{\rho}}{\llen{\rho}}$.
%
We write $\vars{\rho}$ for the set $\{x_{i_0},\dots,x_{i_n}\}$, $\positions{\rho}$ for the set $\{p_0,\dots,p_n\}$, and $\vertices{\rho}$ for the set $\{\xxki{p_0}{i_0},\dots,\xxki{p_n}{i_n}\}$.
We say that $\rho$ is {\em repeating} if $p_0\neq p_n$ and $i_0=i_n$. 
We say that $\rho$ is {\em elementary} if all vertices $\xxki{p_0}{i_0},\dots,\xxki{p_n}{i_n}$ are distinct, with the exception of $\xxki{p_0}{i_0}$ and $\xxki{p_n}{i_n}$, which might be equal.
We say that $\rho$ is {\em essential} if all variables $x_{i_0},\dots,x_{i_n}$ are distinct, with the exception of $x_{i_0}$ and $x_{i_n}$, which might be equal. Clearly, each essential path is also elementary. Note that the length of an essential path is bounded by $N$.
A {\em subpath of $\rho$} is any path of the form $\xxki{p_a}{i_a}\arrow{}{}\dots\arrow{}{}\xxki{p_b}{i_b}$ where $0\leq a\leq b\leq n$. 
We denote by $\overrightarrow{\rho}^{(k)} ~:~ \xxki{p_0+k}{i_0} \arrow{}{} \ldots \arrow{}{} \xxki{p_n+k}{i_n}$ the path obtained by {\em shifting} $\rho$ by $k$, where $k \in \zed$. A path $\rho$ is said to be {\em isomorphic} with another path $\rho'$ if and only if $\rho'=\overrightarrow{\rho}^{(k)}$, for some $k \in \zed$.
Consider a path $\pi = \xxki{q_0}{j_0}\arrow{\beta_0}{}\dots\arrow{\beta_{m-1}}{}\xxki{q_m}{j_m}$. The {\em concatenation} $\rho.\pi$ is defined if $\xxki{p_n}{i_n}=\xxki{q_0}{j_0}$. If $i_n=j_0$, we write $\rho.\pi$ as a shorthand for $\rho.\overrightarrow{\pi}^{(p_n-q_0)}$.
If $\rho$ is repeating and $k\geq 1$, we define the {\em $k$-th power of $\rho$} as the $k$-times concatenation of $\rho$ with itself, e.g. $\rho^3=\rho.\rho.\rho$.
We next define the notion of a {\em compatible} path.
\begin{definition} \label{def:compatible}
%\textbf{(Compatible paths)}
Let $\rho,\rho'$ be paths in $\mathcal{G}_R^n$ for some $n\geq 1$. We say that $\rho'$ is {\em compatible with} $\rho$ (denoted $\rho' \preceq \rho$) if and only if (i) both $\rho$ and $\rho'$ are of the form $\xxki{k}{i}\arrow{}{}\dots\arrow{}{}\xxki{\ell}{j}$ for some $1\leq i,j\leq N$ and $0\leq k,\ell\leq n$, and (ii) $\weight{\rho'}\leq\weight{\rho}$.
\end{definition}
%

%%%%%%%%%%%%%%%%%%%%%%%%%%%%%%%%%%%%%%%%%%%%%%%%%%%%%%%%%%%
\subsubsection{Balanced relations}
%%%%%%%%%%%%%%%%%%%%%%%%%%%%%%%%%%%%%%%%%%%%%%%%%%%%%%%%%%%

%\footnote{Given a formula $\phi$, $\atoms{\phi}$ denotes the set of atomic propositions in $\phi$.}
We say that a difference bounds constraint $R(\x,\x')$ is {\em balanced} whenever $(x-y\leq c)\in\atoms{R}$ if and only if $(x'-y'\leq c)\in\atoms{R}$. Note that the relation $R_b$ (called the {\em balanced closure} of $R$) defined below is balanced:
\[ R_b \stackrel{def}{=} R \wedge \bigwedge_{(x-y\leq c)\in\atoms{R}} x'-y'\leq c \wedge \bigwedge_{(x'-y'\leq c)\in\atoms{R}} x-y\leq c \]
We next show that the computation of the closed form for $R$ can be reduced to the computation of the closed form of its balanced closure:
\begin{proposition} \label{reduction:to:balanced}
Let $R(\x,\x')$ be a DB constraint, $R_b(\x,\x')$ be its balanced closure, and $\widehat{R}_b(\ell,\x,\x')$ be the closed form of $R_b$. Then, $\widehat{R}(k,\x,\x')$ can be defined as:
\[ \bigvee_{i=1}^{2} (k = i \wedge R^i(\x,\x')) \vee \exists \x_1,\x_2 ~.~ k\geq 3 \wedge R(\x,\x_1) \wedge \widehat{R}_b(\ell,\x_1,\x_2)[k-2/\ell] \wedge R(\x_2,\x') \]
\end{proposition}
In the following sections, we study balanced DB relations. Finally, as a consequence of Proposition \ref{reduction:to:balanced}, we show that our results can be generalized to arbitrary DB relations.

%\begin{remark}
%Unlike to techniques based on zigzag automata, we work with vertical edges in constraint graphs, in order to avoid quadratic increase of the number of variables. Balancing is the only extra step needed for easy reasoning about such constraint graphs.
%
%Note that if $\tau=\xxki{k}{i}\arrow{c}{}\xxki{k}{j}$ is a vertical edge in $\mathcal{G}_R^n$ for some $0<k<n$ and $R$ is balanced, then $\tau$ appears both in the $k$-th and the $(k+1)$-th copy of $\mathcal{G}_R$ in $\mathcal{G}_R^n$ and we thus avoid a case split when reasoning about paths.
%\end{remark}

%The reason we work with balanced relations is hinted by the following property. \begin{proposition} Let $R$ be a balanced DB relation, let $n\geq 1$, and let $\rho$ be a path in $\mathcal{G}_R^n$. If $\positions{\rho}=\{a,\dots,b\}$ for some $0\leq a\leq b\leq n$, then there exists an isomorphic path $\rho'$ in $\mathcal{G}_R^{b-a}$. \end{proposition}

%%%%%%%%%%%%%%%%%%%%%%%%%%%%%%%%%%%%%%%%%%%%%%%%%%%%%%%%%%%
%%%%%%%%%%%%%%%%%%%%%%%%%%%%%%%%%%%%%%%%%%%%%%%%%%%%%%%%%%%

%%%%%%%%%%%%%%%%%%%%%%%%%%%%%%%%%%%%%%%%%%%%%%%%%%%%%%%%%%%
%%%%%%%%%%%%%%%%%%%%%%%%%%%%%%%%%%%%%%%%%%%%%%%%%%%%%%%%%%%
\section{Closed Forms for One-directional Difference Bounds Relations}
\label{sec:fw:relations}
We say that a DB constraint $R(\x,\x')$ is {\em one-directional} if it is either (i) a conjunction of the form $\bigwedge_{ij} x_i-x_j'\leq c_{ij}$ ({\em forward} one-directional) or (ii) a conjunction of the form $\bigwedge_{ij} x'_i-x_j\leq c_{ij}$ ({\em backward} one-directional). Clearly, the two cases are dual: $R$ is forward one-directional if and only if its inverse $R^{-1}$ (which can be defined as $R(\x,\x')[\x'/\x,\x/\x']$) is backward one-directional. Consequently, a closed form of $R$ can be directly obtained from a closed form of $R^{-1}$ as:
\[ \widehat{R}(k,\x,\x') \iff \widehat{R^{-1}}(k,\x,\x')[\x/\x',\x'/\x] \]
We can thus consider, without loss of generality, only forward one-directional relations. Let $R$ be such relation. Clearly, $\mathcal{G}_R^n$ contains only forwards edges for all $n\geq 1$. Hence, $\len{\rho}=\llen{\rho}$ for each path $\rho$ in $\mathcal{G}_R^n$ and moreover, $\mathcal{G}_R^n$ contains no cycle and $R^n$ is thus consistent, for all $n\geq 1$. 
Then, by Proposition \ref{computing:powers}, computation of $R^n(\x,\x')$ amounts to computing, for each $1\leq i,j\leq N$, the minimal weight over all paths in $\mathcal{G}_R^n$ of the form $\xxki{0}{i}\arrow{}{}\dots\arrow{}{}\xxki{n}{j}$.
We next show that minimal weight paths have, without loss of generality, regular shape in the sense that they are instances of {\em biquadratic path schemes}:

\begin{definition} \label{def:path:scheme}
If $\sigma,\sigma'$ are paths and $\lambda$ is an empty or an essential repeating path such that $\sigma.\lambda.\sigma'$ is a non-empty path, the expression $\theta=\sigma.\lambda^*.\sigma'$ is called a {\em path scheme}. A path scheme encodes the infinite set of paths $\sem{\theta} = \{\sigma.\lambda^n.\sigma' ~|~ n\geq 0\}$. We say that $\theta$ is {\em biquadratic} if $\len{\sigma.\sigma'} \leq N^4$.
%If $\sigma_1,\dots,\sigma_{k+1}$ are paths and $\lambda_1,\dots,\lambda_k$ are pairwise distinct essential repeating paths for some $k\geq 1$ such that $\sigma_1.\lambda_1\dots\sigma_k.\lambda_k.\sigma_{k+1}$ is a non-empty path, the expression $\theta=\sigma_1.\lambda^*_1.\sigma_2\dots\sigma_k.\lambda^*_k.\sigma_{k+1}$ is called a {\em path scheme}. A path scheme encodes the infinite set of paths $\sem{\theta} = \{\sigma_1.\lambda_1^{n_1}.\sigma_2 \ldots \sigma_k.\lambda_k^{n_k}.\sigma_{k+1} ~|~ n_1,\ldots,n_k \geq 0\}$. We say that $\theta$ is {\em biquadratic} if $\sum_{i=1}^{k+1} \len{\sigma_i} \leq N^4$.
\end{definition}
The following result is a consequence of Lemma 3 in \cite{arxiv-vmcai14}:
\begin{lemma} \label{biquadratic:schemes}
Let $R$ be a one-directional DB relation, let $n\geq 1$, and let $\rho$ be an extremal path in $\mathcal{G}_R^n$. Then, there exists a compatible path $\rho'$ and a biquadratic path scheme $\sigma.\lambda^*.\sigma'$, such that $\rho' \in \sem{\sigma.\lambda^*.\sigma'}$.
\end{lemma}

%Further note that the zigzag automaton of $R$ has at most $N$ control states. We can apply the lemma from VMCAI which tells us that for every minimal run $\rho$ such that $\len{\rho}\geq N^4$, there exists a biquadratic scheme $\sigma.\lambda^*.\sigma$ (i.e. $\len{\sigma.\sigma'}\leq N^4$ and $\len{\lambda}\leq N$) that captures $\rho$. 
%
%We next show that there exists a finite set $\Pi$ of {\em path schemes} such that for any $n\geq 1$, $1\leq i,j\leq N$, and any minimal-weight path from $\xxki{0}{i}$ to $\xxki{n}{j}$ in $\mathcal{G}_R^n$, there exists a path scheme that 
%
%In the following, when $R$ is clear from the context, we implicitly mean $\mathcal{G}_R^n$ (for arbitrary $n\geq 1$) when referring to a path, cycle, etc. 

By Lemma \ref{biquadratic:schemes}, minimal weight paths can be captured by a set $\Pi$ of all biquadratic path schemes. For each such scheme $\sigma.\lambda^*.\sigma' \in \Pi$, we have $\len{\sigma.\sigma'}\leq N^4$ (by Def. \ref{def:path:scheme}) and $\len{\lambda}\leq N$ (since the length of essential paths is bounded by $N$). In the worst case, each vertex of $\mathcal{G}_R^n$ has $N$ successors and hence, there are up to $N^n$ paths in $\mathcal{G}_R^n$ of the form $\xxki{0}{i}\arrow{}{}\dots\arrow{}{}\xxki{n}{j}$, for a fixed $1\leq i,j\leq N$. Consequently, $\card{\Pi}$ is of the order $2^{\mathcal{O}(N)}$. 
We next show that it is sufficient to consider only polynomially many representants from $\Pi$. We first partition $\Pi$ into polynomially many equivalence classes. Each class is determined by (i) first and last variables of $\sigma$, $\lambda$, and $\sigma'$, and by (ii) the length of $\lambda$ and $\sigma.\sigma'$. Formally, the partition is defined as: \[ \Xi \stackrel{def}{=} \{ \Pi_{ijkpq} ~|~ 1\leq i,j,k\leq N, 0\leq p\leq N^4, 0\leq q\leq N, p+q>0 \} \] where each $\Pi_{ijkpq}\subseteq\Pi$ is defined as follows: $\sigma.\lambda^*.\sigma' \in \Pi_{ijkpq}$ if and only if $\sigma,\lambda,\sigma'$ are paths of the form:
\begin{align}
\label{eq:scheme:shape:1}
 \lambda = \xxki{0}{k}\arrow{}{}\dots\arrow{}{}\xxki{p}{k} &
 \\
\label{eq:scheme:shape:2}
\sigma = \xxki{0}{i}\arrow{}{}\dots\arrow{}{}\xxki{r}{k} &,
\hspace{1mm}
\sigma' = \xxki{r}{k}\arrow{}{}\dots\arrow{}{}\xxki{q}{j},
\textrm{ for some } 0\leq r\leq q 
\end{align}
%\begin{equation} \label{eq:scheme:shape}
%\lambda = \xxki{0}{k}\arrow{}{}\dots\arrow{}{}\xxki{p}{k},
%\hspace{1mm}
%\sigma = \xxki{0}{i}\arrow{}{}\dots\arrow{}{}\xxki{r}{k},
%\hspace{1mm}
%\sigma' = \xxki{r}{k}\arrow{}{}\dots\arrow{}{}\xxki{q}{j},
%\textrm{ for some } 0\leq r\leq q
%\end{equation}
%
Intuitively, $p$ ($q$) determines the length of $\lambda$ ($\sigma.\sigma'$) and $k$ determines the variable on which $\lambda$ connects with $\sigma$ and $\sigma'$. Clearly, $\card{\Xi}$ is of the order $\mathcal{O}(N^8)$. 

Let us fix $i,j,k,p,q$ and assume that $\Pi_{ijkpq}\neq\emptyset$. 
It is easy to see that if there exists a path $\lambda$ of the form \eqref{eq:scheme:shape:1}, then there exists one with minimal weight. Similarly, if there exists a path $\sigma.\sigma'$ of the form \eqref{eq:scheme:shape:2}, then there exists one with minimal weight. We define $\theta_{ijkpq}$ as the path scheme $\sigma.\lambda^*.\sigma'$ where $\lambda$ and $\sigma.\sigma'$ are the minimal paths.
%If $p>0$, there clearly exists a minimal $\lambda$ of the form \eqref{eq:scheme:shape}. If $p=0$, we set $\lambda$ to be empty. Similarly, if $q>0$, there exists a minimal $\sigma.\sigma'$ of the form \eqref{eq:scheme:shape}. If $q=0$, we set $\sigma$ and $\sigma'$ to be empty. Then, we define $\theta_{ijkpq} \stackrel{def}{=} \sigma.\lambda.\sigma'$. 
%
It is easy to see that $\theta_{ijkpq}$ is {\em minimal} in $\Pi_{ijkpq}$ in the following sense: $\weight{\sigma.\lambda^n.\sigma'}\leq\weight{\nu.\mu^n.\nu'}$ for each $\nu.\mu^*.\nu'\in\Pi_{ijkpq}$ and each $n\geq 0$. Hence, we can use $\theta_{ijkpq}$ as a representant of $\Pi_{ijkpq}$.
The minimal representants can be computed in polynomial time:
\begin{lemma} \label{scheme:representants:ptime}
The set $\{ \theta_{ijkpq} ~|~ \Pi_{ijkpq}\neq\emptyset \}$ can be computed in PTIME.
\end{lemma}
Next, we fix $1\leq i,j\leq N$ and define:
\[ S_{ij} \stackrel{def}{=} \{ (\len{\sigma.\sigma'},\len{\lambda},\weight{\sigma.\sigma'},\weight{\lambda}) ~|~ \exists k,p,q ~.~ \theta_{ijkpq}=\sigma.\lambda^*.\sigma' \} \]
It follows from the previous arguments that $S_{ij}$ represents all bi-quadratic schemes which capture paths from $\xxki{0}{i}$ to $\xxki{n}{j}$ and moreover, $S_{ij}$ can be computed in polynomial time and its cardinality is polynomial. Then, the closed form of the sequence $\{ \minweight{\xxki{0}{i},\xxki{n}{j},\mathcal{G}_R^n} \}_{n\geq 1}$ can be defined as:
%\begin{equation}\label{eq:cf:forward:0}
\[
\begin{array}{lcl}
\phi_{ij}(n,x_i,x_j') & \iff &
n\geq 1 \wedge
\bigwedge\limits_{(p,q,a,b) \in S_{ij}}
\forall \ell ~.~ (\ell\geq 0 \wedge n=p+q\cdot\ell) \Rightarrow (x_i - x_j' \leq a + b\cdot\ell)
\end{array}
\]
%\end{equation}
%
Intuitively, each conjunct encodes a constraint of one scheme $\sigma.\lambda^*.\sigma'$: whenever the scheme captures a path of length $n$ (i.e. $n=\len{\sigma.\sigma'}+\ell\cdot\len{\lambda} = p+q\cdot\ell$ where $\ell\geq 0$), the difference $x_i-x'_j$ must be upper-bounded by the corresponding weight $\weight{\sigma.\sigma'}+\ell\cdot\weight{\lambda} = a+\ell\cdot b$. Equivalently, we can write:
\begin{equation}\label{eq:cf:forward:one:pair}
\begin{array}{lcl}
\phi_{ij}(n,x_i,x_j') & \iff &
n\geq 1 \wedge
\bigwedge\limits_{(p,q,a,b) \in S_{ij}}
(n\geq p \wedge q ~|~ n-p) \Rightarrow q\cdot(x_i - x_j') \leq q\cdot a + b\cdot(n-p)
\end{array}
\end{equation}
Then, we can define the closed form of $R$ as:
\begin{equation}\label{eq:cf:forward}
\begin{array}{lcl}
\widehat{R}(n,\x,\x') & \iff &
\bigwedge\limits_{1\leq i,j\leq N} \phi_{ij}(n,x_i,x_j') 
\end{array}
\end{equation}
Clearly, $\phi_{ij}(n,x_i,x_j')$ (and hence also $\widehat{R}(n,\x,\x')$) is a formula in the existential fragment of PA and of polynomial size, since $\card{S_{ij}}$ is polynomial. Thus, it follows from Lemma \ref{scheme:representants:ptime} that the whole computation of $\widehat{R}(n,\x,\x')$ is polynomial. %, as summarized by the following theorem.
\begin{theorem} \label{th:cf:forward}
Let $R(\x,\x')$ be a one-directional DB constraint. Then, its closed form $\widehat{R}(n,\x,\x')$ can be computed in PTIME as a formula in the existential fragment of PA.
\end{theorem}

%%%%%%%%%%%%%%%%%%%%%%%%%%%%%%%%%%%%%%%%%%%%%%%%%%%%%%%%%%%
%%%%%%%%%%%%%%%%%%%%%%%%%%%%%%%%%%%%%%%%%%%%%%%%%%%%%%%%%%%

%%%%%%%%%%%%%%%%%%%%%%%%%%%%%%%%%%%%%%%%%%%%%%%%%%%%%%%%%%%
%%%%%%%%%%%%%%%%%%%%%%%%%%%%%%%%%%%%%%%%%%%%%%%%%%%%%%%%%%%
\section{Normalization of Paths in the Unfolded Constraint Graph}
\label{sec:normalization}
%By Proposition \ref{computing:powers}, powers of a DB relation can be defined by computing (minimal) weights of extremal paths. 
In this section, we consider only balanced DB relations and show that every extremal path in an unfolded constraint graph can be {\em normalized}. Intuitively, a path $\rho$ from $\mathcal{G}_R^n$ is normalized if none of its subpaths that traverses only positions in the range $\{N^2,\dots,n-N^2\}$ is a {\em long corner}. 
%it has no {\em long corner} that traverses only positions in the range $\{N^2,\dots,n-N^2\}$.
Informally, a {\em corner} is a vertical path that stays either on the right or on the left side of the initial position. A corner is {\em long} if the distance between its minimal and maximal position exceeds the bound $N^2$. 
\begin{definition} \label{def:corners}
\textbf{(Corners)}
Let $\rho$ be a vertical path of the form $\rho=\xxki{k_0}{i_0}\arrow{}{}\dots\arrow{}{}\xxki{k_m}{i_m}$ for some $m\geq 1$ such that $k_0=k_m$. If $\positions{\rho}=\{k_0,\dots,k_0+d\}$ for some $d\geq 0$, we say that $\rho$ is a {\em right corner of extent} $d$. If $\positions{\rho}=\{k_0-d,\dots,k_0\}$ for some $d\geq 0$, we say that $\rho$ is a {\em left corner of extent} $d$. We say that $\rho$ is a {\em corner} if it is either a left corner or a right corner. We denote the extent of a corner $\rho$ by $\extent{\rho}$. We say that a corner $\rho$ is {\em basic} if $k_0\not\in\{k_1,\dots,k_{m-1}\}$. We say that a corner $\rho$ is {\em long} if $\extent{\rho}>N^2$. We say that $\rho$ is a {\em lb-corner} if it is both long and basic.
\end{definition}
E.g., consider vertical paths from Fig. \ref{fig:dbrel}(c), where $\pi_6$ is a right corner, $\pi_7$ is a right basic corner, and $\pi_5$ is not a corner. Both $\pi_6$ and $\pi_7$ are short, since $\extent{\pi_6}=\extent{\pi_7}=2\leq N^2=5^2$.
In the following, $\lbcorners{\rho}$ ($\lcorners{\rho}$, respectively) denotes the set of subpaths of $\rho$ which are lb-corners (long corners, respectively).
It is easy to show that if a path contains no lb-corner, it also contains no long corner.
We are now ready to formalize the notion of a {\em normalized} path.
%d
\begin{definition} \label{def:normalized}
\textbf{(Normalized paths)}
Let $n\geq 1$ and let $\rho$ be an extremal path in $\mathcal{G}_R^n$. We say that $\rho$ is {\em normalized} if none of its subpaths $\theta$ such that $\positions{\theta}\subseteq\{N^2,\dots,n-N^2\}$ is a long corner.
\end{definition}
E.g., the path in Fig. \ref{figure:segment:transformation}(a) is not normalized, due to the long corner $\theta$.

\subsubsection{Normalization}

We next give a high level idea of normalization. 
Given an integer $n\geq 1$ and an extremal path $\rho_1$ from $\mathcal{G}_R^n$, we construct a finite sequence $\{\rho_k\}_{k=1}^{m}$ of paths from $\mathcal{G}_R^n$ for some $m\geq 1$ such that $\rho_m$ is normalized and $\rho_{k+1}$ is compatible with $\rho_k$, for each $1\leq k\leq m-1$.
%, in the following sense.
By transitivity, we have that $\rho_m$ is compatible with $\rho_1$.
For each $1\leq k<m$, the path $\rho_{k+1}$ is obtained from $\rho_k$ by {\em substituting} some of its subpaths with a compatible path.
\begin{definition} \label{def:substitution}
\textbf{(Substitution)}
If $\rho.\theta.\rho'$ and $\theta'$ are paths in $\mathcal{G}_R^n$ such that $\theta'\preceq\theta$, the {\em substitution of $\theta$ in $\rho.\theta.\rho'$ with $\theta'$} is defined as $(\rho.\theta.\rho')[\theta'/\theta] \stackrel{def}{=} \rho.\theta'.\rho'$.
\end{definition}
The subpaths of $\rho_1,\dots,\rho_{m-1}$ that are substituted are certain paths called {\em segments}:
Informally, {\em segments} of $\rho$ are the unique subpaths of $\rho$ that traverse only positions $\{p,\dots,q\}$ for some fixed parameters $p\leq q$.
\begin{definition} \label{def:segments}
\textbf{(Path segments)} 
Let $n\geq 1$ and let $\rho$ be a path in $\mathcal{G}_R^n$. 
Let $0\leq p\leq q\leq n$ be integers and let $\mathcal{H}$ be the (unique) subgraph of $\mathcal{G}_R^n$ obtained by removing every edge $\tau$ such that $\positions{\tau}\subseteq\{p,\dots,q\}$. We define $\segments{\rho,p,q}$ to be the (unique) sequence $\xi_1,\dots,\xi_m$, for some $m\geq 0$, such that
\begin{itemize}
 \item $\xi_i$ is a subpath of $\rho$ such that $\positions{\xi_i}\subseteq\{p,\dots,q\}$, for each $1\leq i\leq m$
 \item there exist (unique) paths $\sigma_1,\dots,\sigma_{m+1}$ in $\mathcal{H}$ such that $\rho = \sigma_1.\xi_1\dots\sigma_m.\xi_m.\sigma_{m+1}$
\end{itemize}
\end{definition}
In the rest of this paper, we write $\segments{\rho}$ as a shorthand for $\segments{\rho,N^2,n-N^2}$. If $\xi\in\segments{\rho}$, we say that $\xi$ is a {\em segment} of $\rho$.
It is easy to verify that each segment of an extremal path $\rho$ in $\mathcal{G}_R^n$ is of the form $\xi=\xxki{p}{i}\arrow{}{}\dots\arrow{}{}\xxki{q}{j}$ for some $1\leq i,j\leq N$ and $p,q\in\{N^2,n-N^2\}$.
As an example, consider the path in Fig. \ref{figure:segment:transformation}(a), which has one segment $\gamma.\theta.\gamma'$,
The next proposition allows us to use an alternative characterization of normalized paths:
\begin{proposition} \label{normalized:characterization}
Let $n\geq 1$ and let $\rho$ be an extremal path in $\mathcal{G}_R^n$. Then, $\rho$ is {\em normalized} if and only if $\lcorners{\xi}=\emptyset$ for each $\xi\in\segments{\rho}$.
\end{proposition}

\subsubsection{Termination argument}

We argue that the sequence $\{\rho_k\}_{k=1}^{m}$ is finite, by tracking, for each segment, the distance of the {\em first lb-corner} from the end of the segment:
%
%The following proposition can be applied to find the (unique) first lb-corner of a path.
\begin{proposition} \label{unique:first:corner}
\textbf{(Finding the first lb-corner)}
Let $\rho$ be a path such that $\lbcorners{\rho}\neq\emptyset$. Then, $\rho$ has subpaths $\rho_1,\theta,\rho_2$ such that $\rho=\rho_1.\theta.\rho_2$, $\lbcorners{\rho_1.\theta}=\{\theta\}$, and $\extent{\theta}=N^2+1$. The corner $\theta$ is called the {\em first lb-corner of $\rho$}.
\end{proposition}
%
%In the following, the corner $\theta$ from Proposition \ref{unique:first:corner} is called the {\em first lb-corner of $\rho$}.
E.g., $\theta$ is the first lb-corner of the path $\sigma_1.\gamma.\theta.\gamma'.\sigma_2$ in Figure \ref{figure:segment:transformation}(a). We define $\lbsegments{\rho}$ to be the subsequence of $\segments{\rho}$ obtained by erasing every segment $\xi$ such that $\lbcorners{\xi}=\emptyset$. For each $1\leq k<m$, we guarantee that if $\lbsegments{\rho_k}=\xi_1,\dots,\xi_a$ for some $a\geq 1$, then either
\begin{equation}\label{eq:termination:prop}
\begin{array}{llll}
 (1) & \lbsegments{\rho_{k+1}}=\xi_2,\dots,\xi_a \textrm{, or} \\
 (2) & \lbsegments{\rho_{k+1}}=\zeta,\xi_2,\dots,\xi_a \textrm{ for some $\zeta$}
\end{array}
\end{equation}
and moreover, in case (2), $\xi_1$ and $\zeta$ are paths such that
\begin{itemize}
 \item $\xi_1=\gamma_1.\theta_1.\gamma_1'$ for some paths $\gamma_1,\theta_1,\gamma_1'$ and $\theta_1$ is the first lb-corner of $\xi_1$,
 \item $\zeta=\gamma_2.\theta_2.\gamma_2'$ for some paths $\gamma_2,\theta_2,\gamma_2'$ and $\theta_2$ is the first lb-corner of $\zeta$, and
 \item $\len{\gamma_2'}<\len{\gamma_1'}$
\end{itemize}
Intuitively, $\rho_k$ and $\rho_{k+1}$ have the same segments with long corners, with the exception of one segment $\xi_1$, which is either eliminated (case 1), or replaced with another segment $\zeta$ (case 2) such that the length of the unique suffix $\gamma_2'$ of $\zeta$ after its first corner is strictly smaller than the length of the unique suffix $\gamma_1'$ of $\xi_1$ after its first corner. Hence, the number of consecutive applications of the case 2 is bounded by $\len{\xi_1}$. Clearly, if $\len{\xi_1}=0$, then only case 1 may happen, which decreases the number of segments with long corners, and therefore guarantees termination.

\subsubsection{Transforming segments with long corners}

\begin{figure}[t]
\begin{center}
\begin{tabular}{ccc}
\mbox{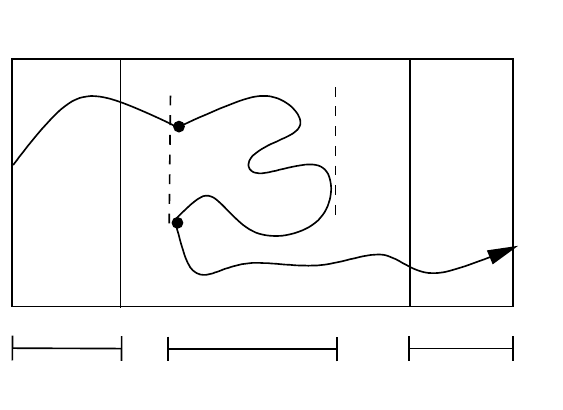}
&
\hspace{5mm}\mbox{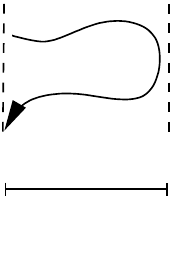}
&
\mbox{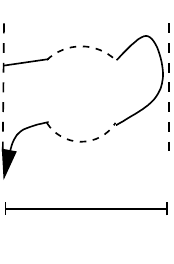}
\\
(a) & \hspace{3mm}(b) & (c)
\\[3mm]
\mbox{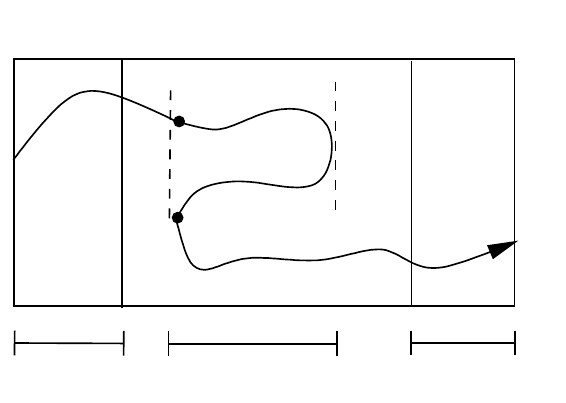}
&
\multicolumn{2}{c}{\mbox{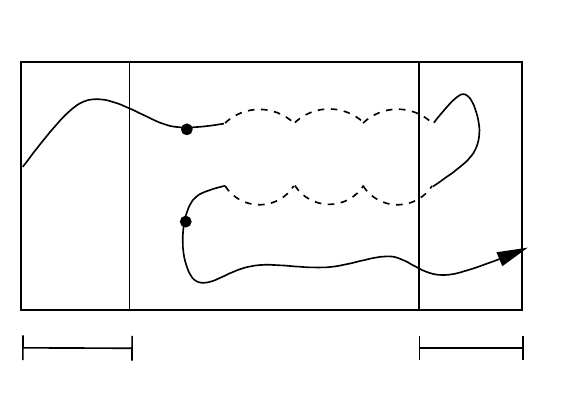}}
\\
(d) & \multicolumn{2}{c}{(e)}
\end{tabular}
\caption{Transformation of a segment with long corners}
\label{figure:segment:transformation}
\end{center}
\vspace{-5mm}
\end{figure}

Let $n\geq 1$, $\rho$ be an extremal path in $\mathcal{G}_R^n$, and let $\xi\in\lbsegments{\rho}$. 
We show how to construct a path $\rho'$ that is compatible with $\rho$ and moreover satisfies the termination properties from \eqref{eq:termination:prop}.
By Proposition \ref{unique:first:corner}, there exists a unique corner $\theta$ such that $\xi=\gamma.\theta.\gamma'$, $\lbcorners{\gamma.\theta}=\{\theta\}$, and $\extent{\theta}=N^2+1$, for some paths $\gamma,\gamma'$. Fig. \ref{figure:segment:transformation}(a) depicts such situation.
Suppose that $\xi$ starts at position $N^2$ and ends at position $n-N^2$ (the other three cases are symmetric). Then, it is not difficult to show that $\theta$ is a right corner.
The following lemma states a key result which allows us to either shorten or decompose the corner $\theta$.
\begin{lemma} \label{corner:decomposition:lemma:merged}
\textbf{(Corner shortening / decomposition)}
Let $\rho$ be a right (left) lb-corner such that $\extent{\rho}=N^2+1$. Then, there exists a compatible right (left) lb-corner $\rho'$ such that either (i) $\extent{\rho'}\leq N^2$ or (ii) $\extent{\rho'}=N^2+1$ and $\rho'$ has subpaths $\eta,\mu,\tau,\mu',\eta'$ such that $\rho'=\eta.\mu.\tau.\mu'.\eta'$, $\mu$ is a forward (backward) repeating path, $\mu'$ is a backward (forward) repeating path, 
%$\eta$ is a forward (backward) path, $\eta'$ is a backward (forward) path, 
$\tau$ is a right (left) corner, $\llen{\eta}=\llen{\eta'}$, $1\leq \llen{\mu}=\llen{\mu'}\leq N^2$, and $\weight{\mu}+\weight{\mu'}<0$. Moreover, for all $k\geq 0$, $\eta.\mu^k.\tau.{\mu'}^k.\eta'$ is a right (left) corner and $\lbcorners{\eta.\mu^k}=\lbcorners{{\mu'}^k.\eta'}=\emptyset$.
\end{lemma}
Let $\theta'$ be the corner obtained by applying Lemma \ref{corner:decomposition:lemma:merged}. Fig. \ref{figure:segment:transformation}(b) depicts the case (i) and Fig. \ref{figure:segment:transformation}(c) the case (ii). 
%
%We next give an intuition about how Lemma \ref{corner:decomposition:lemma:merged} is applied 
%Let $\xi$ be a segment of $\rho$ such that $\lbcorners{\xi}\neq\emptyset$. 

First, suppose that the case (i) of Lemma \ref{corner:decomposition:lemma:merged} applies. Then, one can define
\[ \zeta \stackrel{def}{=} \xi[\theta'/\theta] = (\gamma.\theta.\gamma')[\theta'/\theta] = \gamma.\theta'.\gamma' \] 
(see Fig. \ref{figure:segment:transformation}(d)) and prove that $\lbcorners{\gamma.\theta'}=\emptyset$, by using the fact from Lemma \ref{corner:decomposition:lemma:merged} that $\extent{\theta'}\leq N^2$. If $\lbcorners{\zeta}=\emptyset$, then the case 1 in \eqref{eq:termination:prop} applies. If $\lbcorners{\zeta}\neq\emptyset$, one can infer from $\lbcorners{\gamma.\theta'}=\emptyset$ that the first corner of $\zeta$ involves at least one edge of $\gamma'$ and hence, that the distance of the first lb-corner in $\zeta$ from the end of $\zeta$ strictly decreases, i.e. that the case 2 in \eqref{eq:termination:prop} applies. Hence, the termination property is preserved. We have $\theta'\preceq\theta$, by Lemma \ref{corner:decomposition:lemma:merged}. Consequently, $\zeta\preceq\xi$ and we can define $\rho'\stackrel{def}{=}\rho[\zeta/\xi]$ and see that also $\rho'\preceq\rho$.

Second, suppose that the case (ii) of Lemma \ref{corner:decomposition:lemma:merged} applies. Let $\theta'=\eta.\mu.\tau.\mu'.\eta'$ be the decomposition of $\theta'$ given by Lemma \ref{corner:decomposition:lemma:merged}. Note that $\mu,\mu'$ are repeating and have the same relative length and opposite directions. Hence, we can define the following path ($\ell\geq 1$ is a parameter):
\[ \overline{\theta} \stackrel{def}{=} \eta.\mu^\ell.\tau.{\mu'}^\ell.\eta' \]
We define $\rho'\stackrel{def}{=}\rho[\overline{\theta}/\theta']$. 
By Lemma \ref{corner:decomposition:lemma:merged}, $\weight{\mu}+\weight{\mu'}<0$. Consequently, $\weight{\overline{\theta}}\leq\weight{\theta'}$ for any $\ell\geq 1$ and hence, $\rho'$ is compatible with $\rho$, i.e. $\rho'\preceq\rho$. 
By Lemma \ref{corner:decomposition:lemma:merged}, $1\leq\llen{\mu}=\llen{\mu'}\leq N^2$ and hence, one can choose $\ell$ sufficiently high and make the path $\overline{\theta}$ reach a position in the range $\{n-N^2+1,\dots,n\}$, formally: $n-N^2+1\in\positions{\overline{\theta}}$. See Fig. \ref{figure:segment:transformation}(e) for an illustration. 
%One can also prove that $\overline{\theta}$ is a path in $\mathcal{G}_R^n$ for this choice of $\ell$. Then, essentially, $n-N^2+1\in\positions{\overline{\theta}}$ entails that 
Thus, the segment $\xi$ in $\rho$ is replaced by two segments $\zeta',\zeta$ in $\rho'$. Intuitively, $\zeta'$ has the subpath $\gamma.\eta.\mu^{\ell-1}$ and $\zeta$ has the subpath ${\mu'}^{\ell-1}.\eta'.\gamma'$. Next, we can apply the following property, which is by Lemma \ref{corner:decomposition:lemma:merged}:
\[ \lbcorners{\eta.\mu^k}=\lbcorners{{\mu'}^k.\eta'}=\emptyset \]
to prove that $\lbcorners{\zeta'}=\emptyset$ and $\lbcorners{{\mu'}^{\ell-1}.\eta'}=\emptyset$. The former implies that $\zeta'$ is a segment with no long corners. The latter can then be used to prove that $\zeta$ has the same properties as $\zeta$ in the previous paragraph (for case (i)), i.e. that the termination properties are satisfied in this case as well.

We can thus conclude that every extremal path can be normalized.
\begin{theorem} \label{normalization}
Let $R(\x,\x')$ be a balanced DB constraint, let $n\geq 1$ be an integer, and let $\rho$ be a path between extremal vertices of $\mathcal{G}_R^n$. Then, there exists a normalized path $\rho'$ such that $\rho'\preceq\rho$.
\end{theorem}
\if

%%%%%%%%%%%%%%%%%%%%%%%%%%%%%%%%%%%%%%%%%%%%%%%%%%%%%%%%%%%
%%%%%%%%%%%%%%%%%%%%%%%%%%%%%%%%%%%%%%%%%%%%%%%%%%%%%%%%%%%

%%%%%%%%%%%%%%%%%%%%%%%%%%%%%%%%%%%%%%%%%%%%%%%%%%%%%%%%%%%
%%%%%%%%%%%%%%%%%%%%%%%%%%%%%%%%%%%%%%%%%%%%%%%%%%%%%%%%%%%
\section{Closed Forms for Difference Bounds Relations}
\label{sec:closed:form:db}
By Lemma \ref{computing:powers}, relation $R^n$ is consistent if $\mathcal{G}_R^n$ contains no extremal cycle with negative weight
%Note that for every cycle $\rho$ in $\mathcal{G}_R^n$, there exists an isomorphic cycle $\rho'$ such that $0\in\positions{\rho}$, i.e. $\rho'$ is an extremal path. Moreover, by Lemma \ref{computing:powers}, 
and moreover, consistent relation $R^n$ can be defined as a conjunction of constraints each of which corresponds to a minimal extremal path. Hence, proving that a formula $\phi(\x,\x')$ defines $R^n$ amounts to showing that $\phi(\x,\x')$ implies only those constraints represented by extremal paths in $\mathcal{G}_R^n$. Consequently, a closed form $\widehat{R}(k,\x,\x')$ must satisfy the above for each $k\geq 1$. 
In this section, we show how to define such formula, in several steps. First, we strengthen the relation $R$ in a way that enables us to shortcut every short corner with a single vertical edge (Section \ref{sec:shortcutting:short:corners}). Second, we define a formula that encodes paths that do not contain long corners (Section \ref{sec:encoding:for:short:corners}). Third, we generalize this encoding to extremal paths (Section \ref{sec:encoding:for:normalized:paths}), by exploiting the fact that such paths can be decomposed into segments according to Def. \ref{def:segments} and that segments of extremal normalized paths contain no long corners. Finally, we show how the formula that encodes extremal paths can be used to define a closed form (Section \ref{sec:precise:closed:form}).
%, by relying on the fact that it is sufficient to consider only normalized paths.

%%%%%%%%%%%%%%%%%%%%%%%%%%%%%%%%%%%%%%%%%%%%%%%%%%%%%%%%%%%
%%%%%%%%%%%%%%%%%%%%%%%%%%%%%%%%%%%%%%%%%%%%%%%%%%%%%%%%%%%
\subsection{Shortcutting Short Corners}\label{sec:shortcutting:short:corners}
%%%%%%%%%%%%%%%%%%%%%%%%%%%%%%%%%%%%%%%%%%%%%%%%%%%%%%%%%%%
%%%%%%%%%%%%%%%%%%%%%%%%%%%%%%%%%%%%%%%%%%%%%%%%%%%%%%%%%%%

Consider the strengthened relation $R_s$ in Fig. \ref{figure:corner:shortcut}(a). We prove that for each $n\geq 1$, each short corner in $\mathcal{G}_R^n$ has a compatible vertical edge in $\mathcal{G}_{R_s}^n$ (see Fig. \ref{figure:corner:shortcut}(b-c)).
\begin{figure}[htbp]
%\vspace{-5mm}
\begin{center}
\begin{tabular}{ccc}
\mbox{\begin{minipage}{5.5cm}
\[\begin{array}{l}
R_s(\x,\x') \stackrel{def}{=} R(\x,\x') \wedge S_{fw}(\x) \wedge S_{bw}(\x') \\
\textrm{where } \begin{array}{lcl}
S_{fw}(\x) & \stackrel{def}{=} & \exists \x' ~.~ R^{N^2}(\x,\x') \\
S_{bw}(\x') & \stackrel{def}{=} & \exists \x ~.~ R^{N^2}(\x,\x')
\end{array}
\end{array}\]
\end{minipage}}
&
\mbox{\begin{minipage}{2.9cm}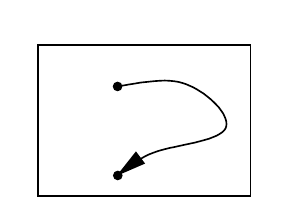\end{minipage}}
&
\mbox{\begin{minipage}{2.9cm}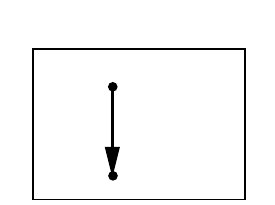\end{minipage}}
\\[10mm]
(a) Strengthened relation $R_s$ & (b) A corner in $\mathcal{G}_{R}^n$ & (c) A vert. edge in $\mathcal{G}_{R_s}^n$
\end{tabular}
\caption{Shortcutting a short corner by strengthening a relation}
\label{figure:corner:shortcut}
\end{center}
\vspace{-5mm}
\end{figure}
\begin{proposition} \label{short:corner:replacement}
\textbf{(Eliminating Short Corners)}
Let $R(\x,\x')$ be a balanced DB constraint, let $n\geq 1$ be an integer and let $\theta$ be a short corner in $\mathcal{G}_R^n$ of the form $\xxki{k}{i}\arrow{}{}\dots\arrow{}{}\xxki{k}{j}$. Then:
\[\begin{array}{lcl}
  R_s(\x,\x') \Rightarrow S_{fw}(\x) \Rightarrow x_i-x_j\leq\weight{\theta} &\hspace{10mm}& \textrm{(if $\theta$ is right)} \\
  R_s(\x,\x') \Rightarrow S'_{bw}(\x') \Rightarrow x'_i-x'_j\leq\weight{\theta} && \textrm{(if $\theta$ is left)}
\end{array}\]
Consequently, there is a compatible vertical edge $\xxki{k}{i}\arrow{c}{}\xxki{k}{j}$ in $\mathcal{G}_{R_s}^n$, for some $c\leq\weight{\theta}$.
\end{proposition}
The intuition is that if we view the above short right corner $\theta$ as an extremal path in $\mathcal{G}_R^{N^2}$ that starts at position $0$, then we have, by Lemma \ref{computing:powers}, that $R^{N^2}(\x,\x') \Rightarrow x_i-x_j\leq\weight{\theta}$, and hence the first implication in Proposition \ref{short:corner:replacement} holds, by the definition of $S_{fw}$ and $R_s$.

%%%%%%%%%%%%%%%%%%%%%%%%%%%%%%%%%%%%%%%%%%%%%%%%%%%%%%%%%%%
%%%%%%%%%%%%%%%%%%%%%%%%%%%%%%%%%%%%%%%%%%%%%%%%%%%%%%%%%%%
\subsection{Encoding Paths without Long Corners}\label{sec:encoding:for:short:corners}
%%%%%%%%%%%%%%%%%%%%%%%%%%%%%%%%%%%%%%%%%%%%%%%%%%%%%%%%%%%
%%%%%%%%%%%%%%%%%%%%%%%%%%%%%%%%%%%%%%%%%%%%%%%%%%%%%%%%%%%

The strengthening from Section \ref{sec:shortcutting:short:corners} can be used to {\em straighten} paths which have only short corners. Informally, a straightened path is either (i) a sequence of forward edges, or (ii) a sequence of backward edges, or (iii) a single vertical edge. Let $\xi$ be an extremal path in $\mathcal{G}_{R}^n$ such that $\lcorners{\xi}=\emptyset$. First, suppose that $\xi$ is forward. 
\begin{figure}[b]
\vspace{-5mm}
%\begin{wrapfigure}{r}{8cm}
%\vspace{-11mm}
\begin{center}
\begin{tabular}{cccc}
\mbox{\begin{minipage}{4.0cm}
\[\begin{array}{lcl} 
R_{fw} & \stackrel{def}{=} \bigwedge \{ & x-y'\leq c ~|~ \\ && R_s \Rightarrow x-y'\leq c \} \\
R_{bw} & \stackrel{def}{=} \bigwedge \{ & x'-y\leq c ~|~ \\ && R_s \Rightarrow x'-y\leq c \}
\end{array}\]
\end{minipage}}
&
\mbox{\begin{minipage}{2.5cm}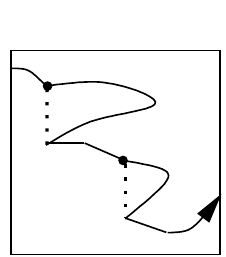\end{minipage}}
&
\mbox{\begin{minipage}{2.5cm}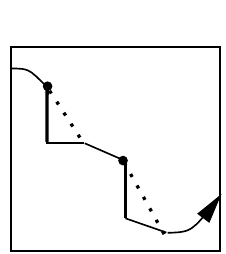\end{minipage}}
&
\mbox{\begin{minipage}{2.5cm}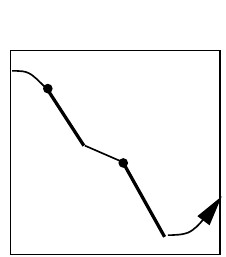\end{minipage}}
\\[12mm]
(a) & (b) A path in $\mathcal{G}_{R}^n$ & (c) A path in $\mathcal{G}_{R_s}^n$ & (d) A path in $\mathcal{G}_{R_{fw}}^n$
\end{tabular}
\end{center}
\vspace{-5mm}
\caption{Path straightening}
\label{figure:straightening}
%\end{figure}
%\vspace{-8mm}
%\end{wrapfigure}
\vspace{-5mm}
\end{figure}
Then, $\xi$ can viewed as a sequence of forward edges and right corners in $\mathcal{G}_{R}^n$ (Fig. \ref{figure:straightening}(b)). By Proposition \ref{short:corner:replacement}, each corner can be shortcut by a vertical edge, and hence we obtain an equivalent path $\xi'$ in $\mathcal{G}_{R_s}^n$ which is a sequence of forward and vertical edges in $\mathcal{G}_{R_{s}}^n$ (Fig. \ref{figure:straightening}(c)). Then, every subpath of $\xi'$ of the form $(vertical\mbox{-}edge)^*.fw\mbox{-}edge$ can be replaced by a (transitively) implied forward edge in $\mathcal{G}_{R_{fw}}^n$ where $R_{fw}$ is defined in Fig. \ref{figure:straightening}(a).
%
%\[\begin{array}{lclclcl} 
%R_{fw} & \stackrel{def}{=} & \bigwedge \{ x-y'\leq c ~|~ R_s \Rightarrow x-y'\leq c \} &\hspace{5mm}&
%R_{bw} & \stackrel{def}{=} & \bigwedge \{ x'-y\leq c ~|~ R_s \Rightarrow x'-y\leq c \}
%\end{array}\]
and thus obtaining an equivalent path $\xi''$ in $\mathcal{G}_{R_{fw}}^n$ that contains only forward edges (Fig. \ref{figure:straightening}(d)). Then, $\xi''$ is encoded by $\widehat{R}_{fw}(\ell,\x,\x')[n/\ell]$ and hence also in $\phi(\ell,\x,\x')[n/\ell]$ defined as:
\begin{equation} \label{eq:enc:segment:balanced}
\phi(\ell,\x,\x') \iff \widehat{R}_{fw}(\ell,\x,\x') \wedge \widehat{R}_{bw}(\ell,\x,\x') \wedge S_{fw}(\x) \wedge S_{bw}(\x')
\end{equation}
If $\xi$ is an extremal right corner, then it is encoded by $S_{fw}(\x)$ (by Proposition \ref{short:corner:replacement}) and hence also by $\phi(\ell,\x,\x')$ (since $S_{fw}$ is its conjunct). The other cases (backward extremal path, extremal left corner) are symmetric. 
%
%Intuitively, $R_{fw}$ preserves only constraints between unprimed and primed variables.
%
Hence, $\phi(\ell,\x,\x')$ encodes all extremal paths in $\mathcal{G}_R^n$ that have no long corners, in the following sense:
\begin{proposition} \label{encoding:of:segments}
\textbf{(Encoding of paths without long corners)}
Let $R(\x,\x')$ be a balanced DB constraint, let $n\geq 1$, and let $\xi$ be an extremal path in $\mathcal{G}_R^n$, i.e. of the form $\xxki{p}{i}\arrow{}{}^+\xxki{q}{j}$ for some $1\leq i,j\leq N$ and $p,q\in\{0,n\}$. If $\lcorners{\xi}=\emptyset$, then:
\[\begin{array}{lclcl}
1. && \phi(\ell,\x,\x')[n/\ell] \Rightarrow x_i-x'_j\leq\weight{\xi} &\hspace{5mm}& \textrm{if $p=0$, $q=n$} \\
2. && \phi(\ell,\x,\x')[n/\ell] \Rightarrow x'_i-x_j\leq\weight{\xi} && \textrm{if $p=n$, $q=0$} \\
3. && \phi(\ell,\x,\x')[n/\ell] \Rightarrow x_i-x_j\leq\weight{\xi} && \textrm{if $p=q=0$} \\
4. && \phi(\ell,\x,\x')[n/\ell] \Rightarrow x'_i-x'_j\leq\weight{\xi} && \textrm{if $p=q=n$}
\end{array}\]
\end{proposition}
%

%%%%%%%%%%%%%%%%%%%%%%%%%%%%%%%%%%%%%%%%%%%%%%%%%%%%%%%%%%%
%%%%%%%%%%%%%%%%%%%%%%%%%%%%%%%%%%%%%%%%%%%%%%%%%%%%%%%%%%%
\subsection{Encoding Extremal Paths}\label{sec:encoding:for:normalized:paths}
%%%%%%%%%%%%%%%%%%%%%%%%%%%%%%%%%%%%%%%%%%%%%%%%%%%%%%%%%%%
%%%%%%%%%%%%%%%%%%%%%%%%%%%%%%%%%%%%%%%%%%%%%%%%%%%%%%%%%%%

Consider the following formula (let $\y$ and $\z$ be fresh copies of variables in $\x$):
\begin{equation} \label{eq:enc:norm:path:balanced}
 \psi(\ell,\x,\x') \iff \exists \y,\z ~.~ R^{N^2}(\x,\y) \wedge \phi(\ell,\y,\z) \wedge R^{N^2}(\z,\x')
\end{equation}
We prove that for each $n\geq 1$, the formula $\psi(\ell,\x,\x')$ encodes every extremal path in $\mathcal{G}_R^n$, in the following sense.
\begin{proposition} \label{encoding:of:extremal:paths}
\textbf{(Encoding of extremal paths)}
Let $R(\x,\x')$ be a balanced DB constraint, let $n\geq 1$, and let $\rho$ be an extremal normalized path in $\mathcal{G}_R^{2N^2+n}$. Then, $\rho$ is of the form $\xxki{p}{i}\arrow{}{}^+\xxki{q}{j}$ for some $1\leq i,j\leq N$ and $p,q\in\{0,2N^2+n\}$ and moreover:
\[\begin{array}{lclcl}
1. && \psi(\ell,\x,\x')[n/\ell] \Rightarrow x_i-x'_j\leq\weight{\rho} &\hspace{5mm}& \textrm{if $p=0$, $q=2N^2+n$} \\
2. && \psi(\ell,\x,\x')[n/\ell] \Rightarrow x'_i-x_j\leq\weight{\rho} && \textrm{if $p=2N^2+n$, $q=0$} \\
3. && \psi(\ell,\x,\x')[n/\ell] \Rightarrow x_i-x_j\leq\weight{\rho} && \textrm{if $p=q=0$} \\
4. && \psi(\ell,\x,\x')[n/\ell] \Rightarrow x'_i-x'_j\leq\weight{\rho} && \textrm{if $p=q=2N^2+n$}
\end{array}\]
\end{proposition}
The intuition behind the encoding is as follows. Let $\rho$ be an extremal normalized path and let $\rho=\sigma_1.\xi_1\dots\sigma_m.\xi_m.\sigma_{m+1}$ be its decomposition according to Def. \ref{def:segments}. By Prop. \ref{normalized:characterization}, $\lcorners{\xi_i} = \emptyset$ for each $1\leq i\leq m$, and hence, by Prop. \ref{encoding:of:segments}, $\xi_i$ is encoded by $\phi(\ell,\y,\z)$. For each $1\leq i\leq m+1$, we have that $\sigma_i$ is encoded in $R^{N^2}(\x,\y)$ or in $R^{N^2}(\z,\x')$. Then, since $\rho=\sigma_1.\xi_1\dots\sigma_m.\xi_m.\sigma_{m+1}$, one can show, by transitivity, that \eqref{eq:enc:norm:path:balanced} encodes $\rho$. E.g., consider the path $\rho=\sigma_1.\gamma.\theta'.\gamma'.\sigma_2$ in Figure \ref{figure:segment:transformation}(d) and denote $\xi_1=\gamma.\theta'.\gamma'$. Supposing $\rho$ is normalized, we have:
\[
\left(\begin{array}{lcl}
R^{N^2}(\x,\y) & \Rightarrow &
x_{i_1}-y_{i_2}\leq\weight{\sigma_1} 
\\
\phi(\ell,\y,\z)[n/\ell] & \Rightarrow & y_{i_2}-z_{i_3}\leq\weight{\xi_1} 
\\
R^{N^2}(\z,\x') & \Rightarrow & z_{i_3}-x'_{i_4}\leq\weight{\sigma_2} 
\end{array}\right)
\Rightarrow 
\left(\begin{array}{l}
\psi(\ell,\x,\x')[n/\ell] \Rightarrow \\
x_{i_1}-x'_{i_4}\leq\weight{\sigma_1.\xi_1.\sigma_2}=\weight{\rho} 
\end{array}\right)
\]

%%%%%%%%%%%%%%%%%%%%%%%%%%%%%%%%%%%%%%%%%%%%%%%%%%%%%%%%%%%
%%%%%%%%%%%%%%%%%%%%%%%%%%%%%%%%%%%%%%%%%%%%%%%%%%%%%%%%%%%
\subsection{Defining the Closed Form}\label{sec:precise:closed:form}
%%%%%%%%%%%%%%%%%%%%%%%%%%%%%%%%%%%%%%%%%%%%%%%%%%%%%%%%%%%
%%%%%%%%%%%%%%%%%%%%%%%%%%%%%%%%%%%%%%%%%%%%%%%%%%%%%%%%%%%

We finally prove that the formula $\widehat{R}(k,\x,\x')$ defined felow is a closed form of $R$:
\begin{equation} \label{eq:balanced:closed:form}
\widehat{R}(k,\x,\x') \iff \bigvee_{i=1}^{2N^2} (k=i \wedge R^i(\x,\x')) \vee \exists \ell\geq 1 ~.~ k=2N^2+\ell \wedge \psi(\ell,\x,\x')
\end{equation}
Note that $R_{fw}$ and $R_{bw}$ are one-directional DB relations (see Section \ref{sec:fw:relations}). Clearly, $S_{fw}$, $S_{fw}$, $R_s$, $R_{fw}$, and $R_{bw}$ are PTIME-computable DB constraints, by Lemma \ref{computing:powers} and Proposition \ref{dbm:exist:elim}. Since $\widehat{R}_{fw}(\ell,\x,\x')$ and $\widehat{R}_{bw}(\ell,\x,\x')$ are PTIME-computable formulas in the existential fragment of PA, by Theorem \ref{th:cf:forward}, so is the formula $\phi(\ell,\x,\x')$ in \eqref{eq:enc:segment:balanced}, and hence also $\widehat{R}(k,\x,\x')$ in \eqref{eq:balanced:closed:form}. 
\begin{theorem} \label{balanced:closed:form}
Let $R(\x,\x')$ be a balanced DB constraint. Then, \eqref{eq:balanced:closed:form} defines a closed form of $R(\x,\x')$. Moreover, $\widehat{R}(n,\x,\x')$ is a PTIME-computable formula in the existential fragment of PA.
\end{theorem}
By Proposition \ref{reduction:to:balanced}, the result of Theorem \ref{balanced:closed:form} extends to arbitrary DB relation.
\begin{corollary} \label{closed:form}
Let $R(\x,\x')$ be a DB constraint. Then, its closed form is a PTIME-computable formula from the existential fragment of PA.
%$R_b$ its balanced closure, and $\widehat{R_b}(\ell,\x,\x')$ be the closed form of $R_b$. Then, the formula
%\begin{equation} \label{eq:closed:form}
%\widehat{R}(n,\x,\x') \iff \bigvee_{i=1}^{2} (n=i \wedge R^i(\x,\x')) \vee \exists \ell\geq 1,\y,\z ~.~ n=2+\ell \wedge R(\x,\y) \wedge \widehat{R_b}(\ell,\y,\z) \wedge R(\z,\x')
%\end{equation}
\end{corollary}

%%%%%%%%%%%%%%%%%%%%%%%%%%%%%%%%%%%%%%%%%%%%%%%%%%%%%%%%%%%
%%%%%%%%%%%%%%%%%%%%%%%%%%%%%%%%%%%%%%%%%%%%%%%%%%%%%%%%%%%

%%%%%%%%%%%%%%%%%%%%%%%%%%%%%%%%%%%%%%%%%%%%%%%%%%%%%%%%%%%
%%%%%%%%%%%%%%%%%%%%%%%%%%%%%%%%%%%%%%%%%%%%%%%%%%%%%%%%%%%
\section{Octagonal Relations}
\label{sec:octagons}

The class of integer octagonal constraints is defined as follows:
\begin{definition}\label{odbc}
  A~formula $\phi(\vec{x})$ is an \emph{octagonal constraint} if it is
  a~finite conjunction of terms of the form $x_i - x_j \le a_{ij}$,
  $x_i + x_j \le b_{ij}$ or $-x_i - x_j \le c_{ij}$ where $a_{ij},
  b_{ij}, c_{ij} \in \zed$, for all $1 \le i,j\le N$.  A~relation
  $R\subseteq\zedNNval$ is an~{\em octagonal relation} if it can be
  defined by an octagonal constraint $\phi_R(\x,\x')$.
\end{definition}
We represent octagons as difference bounds constraints over the dual
set of variables $\vec{y} = \{y_1,y_2,\ldots,y_{2N}\}$, with the
convention that $y_{2i-1}$ stands for $x_i$ and $y_{2i}$ for $-x_i$,
respectively. For example, the octagonal constraint $x_1+x_2=3$ is
represented as $y_1 - y_4 \leq 3 \wedge y_2 - y_3 \leq -3$. In order
to handle the $\vec{y}$ variables in the following, we define
$\bar{\imath} = i-1$, if $i$ is even, and $\bar{\imath}=i+1$ if $i$ is
odd. Obviously, we have $\bar{\bar{\imath}} = i$, for all $i \in
\nat$. We denote by $\overline{\phi}(\y)$ the difference bounds
constraint over $\y$ that represents $\phi(\x)$:
\begin{definition} \label{def:oct:to:dbc}
  Given an octagonal constraint $\phi(\x)$, $\x=\{x_1,\dots,x_N\}$,
  its {\em difference bounds representation} $\overline{\phi}(\y)$,
  over $\y=\{y_1,\dots,y_{2N}\}$, is a~conjunction of the following
  difference bounds constraints, where $1\leq i,j \leq N$,
  $c\in\zed$.
  \[ \begin{array}{lcl}
    (x_i-x_j\leq c)\in \atoms{\phi} & \Leftrightarrow & (y_{2i-1}-y_{2j-1}\leq c), (y_{2j}-y_{2i}\leq c)\in \atoms{\overline{\phi}} \\
    (-x_i+x_j\leq c)\in \atoms{\phi} & \Leftrightarrow & (y_{2j-1}-y_{2i-1}\leq c), (y_{2i}-y_{2j}\leq c)\in \atoms{\overline{\phi}} \\
    (-x_i-x_j\leq c)\in \atoms{\phi} & \Leftrightarrow & (y_{2i}-y_{2j-1}\leq c), (y_{2j}-y_{2i-1}\leq c)\in \atoms{\overline{\phi}} \\
    (x_i+x_j\leq c)\in \atoms{\phi} & \Leftrightarrow & (y_{2i-1}-y_{2j}\leq c), (y_{2j-1}-y_{2i}\leq c)\in \atoms{\overline{\phi}} 
%    (2x_i\leq c)\in \atoms{\phi} & \Leftrightarrow & (y_{2i-1}-y_{2i}\leq c)\in \atoms{\overline{\phi}} \\
%    (-2x_i\leq c)\in \atoms{\phi} & \Leftrightarrow & (y_{2i}-y_{2i-1}\leq c)\in \atoms{\overline{\phi}}
  \end{array} \]
\end{definition}
The following result has been proved in \cite{arxiv-lmcs}.
\begin{lemma} \label{oct:consistent:powers}
Let $n\geq 1$ and let $R(\x,\x')$ be an octagonal relation. Then, if $R^n(\x,\x')$ is consistent, the following equivalence holds:
\[ R^n(\x,\x') \iff \overline{R}^n(\y,\y')[\sigma]  \textrm{, \hspace{3mm} where \hspace{3mm}} \sigma = [x_i/y_{2i-1},-x_i/y_{2i},x'_i/y'_{2i-1},x'_i/y'_{2i}]_{i=1}^{N} \]
\end{lemma}
Hence, a consistent $n$-th power of $R(\x,\x')$ can be computed by applying the above substitution $\sigma$ on the $n$-th power of $\overline{R}(\y,\y')$.
%An immediate consequence of Lemma \ref{oct:consistent:powers} is that if $R^n(\x,\x')$ is consistent, the $n$-th power of $R$ can be computed from the closed form of $\overline{R}(\y,\y')$:
%\[ R^n(\x,\x') \iff \widehat{\overline{R}}(k,\y,\y')[n/k][\sigma] \]
%

\subsubsection{Checking $*$-consistency}
We say that a relation $R$ is {\em $*$-consistent} if $R^n$ is consistent for each $n\geq 1$. If $R$ is not $*$-consistent, we define the minimal inconsistent power of $R$ as:
\[ K_R \stackrel{def}{=} \min\{ n ~|~ n\geq 1, R^n \textrm { is inconsistent} \} \]
\begin{lemma}\label{min:inconsistent}
Checking $*$-consistency of $R$ and computation of $K_R$ can be done in PTIME.
\end{lemma}

\subsubsection{Closed form}
We prove that the closed form of $R$ can be defined as
\begin{equation}\label{eq:cf:octagons}
\widehat{R}(k,\x,\x') \iff \left\{\begin{array}{lcl}
 \widehat{\overline{R}}(k,\y,\y')[\sigma] &\hspace{6mm}&  \textrm{ if $R$ is $*$-consistent} \\
 \widehat{\overline{R}}(k,\y,\y')[\sigma] \wedge k<K_R &\hspace{6mm}&  \textrm{ otherwise} \\
\end{array}\right.
\end{equation}

\begin{theorem} \label{closed:form:octagons}
Let $R(\x,\x')$ be an octagonal constraint. Then, \eqref{eq:cf:octagons} defines its closed form and moreover, it is a PTIME-computable formula in the existential fragment of PA.
\end{theorem}

%%%%%%%%%%%%%%%%%%%%%%%%%%%%%%%%%%%%%%%%%%%%%%%%%%%%%%%%%%%
%%%%%%%%%%%%%%%%%%%%%%%%%%%%%%%%%%%%%%%%%%%%%%%%%%%%%%%%%%%

%%%%%%%%%%%%%%%%%%%%%%%%%%%%%%%%%%%%%%%%%%%%%%%%%%%%%%%%%%%
%%%%%%%%%%%%%%%%%%%%%%%%%%%%%%%%%%%%%%%%%%%%%%%%%%%%%%%%%%%
\section{Conclusions}
\label{sec:conclusions}
%%%%%%%%%%%%%%%%%%%%%%%%%%%%%%%%%%%%%%%%%%%%%%%%%%%%%%%%%%%
%%%%%%%%%%%%%%%%%%%%%%%%%%%%%%%%%%%%%%%%%%%%%%%%%%%%%%%%%%%

We have presented a method that computes transitive closures of octagonal relations in polynomial time. This result also provides a proof of the fact that transitive closures are expressible in (the existential fragment of) Presburger arithmetic. Consequently, our result also simplifies the proof of NP-completeness of reachability checking for flat counter automata, by allowing a deterministic polynomial time reduction to the satisfiability of QFPA.

\bibliographystyle{plain}
\bibliography{main}

\begin{thebibliography}{10}

\bibitem{tacas09}
M.~Bozga, C.~G\^{\i}rlea, and R.~Iosif.
\newblock Iterating octagons.
\newblock In {\em Proc. of TACAS}, volume 5505 of {\em LNCS}, pages 337--351,
  Berlin, Heidelberg, 2009. Springer Verlag.

\bibitem{cav10}
M.~Bozga, R.~Iosif, and F.~Kone\v{c}n\'{y}.
\newblock Fast acceleration of ultimately periodic relations.
\newblock In {\em Proc. of CAV}, volume 6174 of {\em LNCS}, pages 227--242,
  Berlin, Heidelberg, 2010. Springer Verlag.

\bibitem{arxiv-lmcs}
M.~{Bozga}, R.~{Iosif}, and F.~{Kone\v{c}n\'{y}}.
\newblock Deciding conditional termination.
\newblock Technical Report arXiv 1302.2762, 2013.

\bibitem{arxiv-vmcai14}
M.~{Bozga}, R.~{Iosif}, and F.~{Kone\v{c}n\'{y}}.
\newblock Safety problems are {NP}-complete for flat integer programs with
  octagonal loops.
\newblock Technical Report arXiv 1307.5321, 2013.

\bibitem{vmcai14}
M.~Bozga, R.~Iosif, and F.~Kone\v{c}n{\'y}.
\newblock Safety problems are {NP}-complete for flat integer programs with
  octagonal loops.
\newblock In {\em Proc. of VMCAI}, pages 242--261, 2014.

\bibitem{fundamenta}
M.~Bozga, R.~Iosif, and Y.~Lakhnech.
\newblock Flat parametric counter automata.
\newblock {\em Fundamenta Informaticae}, 91(2):275--303, 2009.

\bibitem{comon-jurski}
H.~Comon and Y.~Jurski.
\newblock Multiple counters automata, safety analysis and presburger
  arithmetic.
\newblock In {\em Proc. of CAV}, volume 1427 of {\em LNCS}, pages 268--279,
  Berlin, Heidelberg, 1998. Springer Verlag.

\bibitem{atva12}
H.~Hojjat, R.~Iosif, F.~Kone\v{c}n{\'y}, V.~Kuncak, and P.~R{\"u}mmer.
\newblock Accelerating interpolants.
\newblock In {\em Proc. of ATVA}, pages 187--202, 2012.

\bibitem{fm12}
H.~Hojjat, F.~Kone\v{c}n{\'y}, F.~Garnier, R.~Iosif, V.~Kuncak, and
  P.~R{\"u}mmer.
\newblock A verification toolkit for numerical transition systems - tool paper.
\newblock In {\em Proc. of FM}, pages 247--251, 2012.

\bibitem{mine-thesis}
A.~Min\'e.
\newblock {\em Weakly Relational Numerical Abstract Domains}.
\newblock 2004.

\bibitem{mine06}
A.~Min\'e.
\newblock The octagon abstract domain.
\newblock {\em Higher-Order and Symbolic Computation}, 19(1):31--100, 2006.

\bibitem{WangIGG05}
C.~Wang, F.~Ivancic, M.~K. Ganai, and A.~Gupta.
\newblock Deciding separation logic formulae by {SAT} and incremental negative
  cycle elimination.
\newblock In {\em LPAR}, pages 322--336, 2005.

\end{thebibliography}

\appendix

\if
\else
%%%%%%%%%%%%%%%%%%%%%%%%%%%%%%%%%%%%%%%%%%%%%%%%%%%%%%%%%%%
%%%%%%%%%%%%%%%%%%%%%%%%%%%%%%%%%%%%%%%%%%%%%%%%%%%%%%%%%%%
\section{A Closed Form Based on Periodic Characterization}
%%%%%%%%%%%%%%%%%%%%%%%%%%%%%%%%%%%%%%%%%%%%%%%%%%%%%%%%%%%
%%%%%%%%%%%%%%%%%%%%%%%%%%%%%%%%%%%%%%%%%%%%%%%%%%%%%%%%%%%

\subsection{Motivation}

The problem with the size of the prefix and the period can be illustrated by the DB relation defined below.
\begin{figure} \label{ex:fw:exponential}
\begin{center}\begin{tabular}{cc}
\begin{minipage}{7cm}
\[ \begin{array}{lcl}
R & \Leftrightarrow & x_1-x_2'\leq 0 \wedge x_2-x_1'\leq 0 \wedge x_4-x_5'\leq 0 ~\wedge \\
&& x_5-x_6'\leq 0 \wedge x_6-x_4'\leq 0 \wedge x_3-x_3'\leq 1 ~\wedge \\
&& x_3-x_2'\leq 50 \wedge x_3-x_4'\leq 70 ~\wedge \\
&& x_2-x_3'\leq 0 \wedge x_4-x_3'\leq 0
\end{array} \]
\end{minipage}
&
\begin{minipage}{3cm}
      \scalebox{0.85}{\begin{tikzpicture}
        \TermGridGenDifferentCap{0.0}{0.9}{2}{0.0}{0.7}{6}{0.5}{0.3}{0}
        \foreach \ii in {1,...,1} {
          \pgfmathtruncatemacro\jj{\ii+1}
          \TermGridEdgeC{\ii}{1}{\jj}{2}{\tiny$0$}{above}
          \TermGridEdgeC{\ii}{2}{\jj}{1}{\tiny$0$}{left}
          \TermGridEdgeC{\ii}{4}{\jj}{5}{\tiny$0$}{below}
          \TermGridEdgeC{\ii}{5}{\jj}{6}{\tiny$0$}{below}
          \TermGridEdgeC{\ii}{6}{\jj}{4}{\tiny$0$}{below}
          \TermGridEdgeC{\ii}{3}{\jj}{3}{\tiny$1$}{above}
          \TermGridEdgeC{\ii}{3}{\jj}{2}{\tiny$50$}{right}
          \TermGridEdgeC{\ii}{2}{\jj}{3}{\tiny$0$}{left}
          \TermGridEdgeC{\ii}{3}{\jj}{4}{\tiny$70$}{right}
          \TermGridEdgeC{\ii}{4}{\jj}{3}{\tiny$0$}{left}
%          \TermGridEdgeC{\jj}{6}{\ii}{6}{\tiny$1$}{very near start,above,yshift=-0.5mm}
        }
      \end{tikzpicture}}
\end{minipage}
\end{tabular}\end{center}
\end{figure}
This relation is of the form $\bigwedge_{ij} x_i-x_j'\leq c_{ij}$ and belongs to a fragment called {\em one-directional} DB relations (studied in Section \ref{sec:fw:relations}).
Note that e.g. the sequence $\{min\_weight(\xxki{0}{3},\xxki{n}{3},\mathcal{G}_{R}^n)\}_{n\geq 1}$ has prefix $70$ (so it depends on the constants in $R$) and period $\llcm(2,3)$ (since $R$ has two permuting blocks of variables $\{x_1,x_2\}$ and $\{x_4,x_5,x_6\}$ with optimal paths). %Note that if $N\geq 1$ is the number of variables, one can construct  that one can 

\fi

%%%%%%%%%%%%%%%%%%%%%%%%%%%%%%%%%%%%%%%%%%%%%%%%%%%%%%%%%%%
%%%%%%%%%%%%%%%%%%%%%%%%%%%%%%%%%%%%%%%%%%%%%%%%%%%%%%%%%%%
\section{Remaining Proofs from Section \ref{sec:dbrel}}
%%%%%%%%%%%%%%%%%%%%%%%%%%%%%%%%%%%%%%%%%%%%%%%%%%%%%%%%%%%
%%%%%%%%%%%%%%%%%%%%%%%%%%%%%%%%%%%%%%%%%%%%%%%%%%%%%%%%%%%

%\proof{
%\bigskip\noindent{\bf Proof of Lemma \ref{computing:powers}:}
%See Lemma 6 in \cite{arxiv-vmcai14}.
%\qed
%}

The following technical proposition states that if $R$ is balanced, paths in $\mathcal{G}_R^n$ can be shifted arbitrarily in $\mathcal{G}_R^n$. Note that this claim doesn't hold if $R$ is not balanced, since e.g. an extremal vertical edge $\rho=\xxki{0}{i}\arrow{}{}\xxki{0}{j}$ in $\mathcal{G}_R^n$ has copies $\overrightarrow{\rho}^{(k)}=\xxki{k}{i}\arrow{}{}\xxki{k}{j}$ for each $k=0,\dots,n-1$, but possibly not for $k=n$.
\begin{proposition}\label{balanced:shifting}
Let $R(\x,\x')$ be a balanced DB constraint, let $n\geq 1$, and let $\rho$ be a path in $\mathcal{G}_R^n$ such that $\positions{\rho}=\{a,\dots,b\}$ for some $0\leq a\leq b\leq n$. Then, $\overrightarrow{\rho}^{(k)}$ is an isomorphic path in $\mathcal{G}_R^n$ such that $\positions{\overrightarrow{\rho}^{(k)}}=\{a+k,\dots,b+k\}$, for each $k\in\{-a,\dots,n-b\}$.
\end{proposition}

%\ifLongVersion\proof{
\bigskip\noindent{\bf Proof of Proposition \ref{reduction:to:balanced}:}
It is easy to verify that $R\circ R_b^n\circ R = R^{2+n}$ for all $n\geq 0$. Then, the claim is a direct consequence of this fact.
\qed
%}\fi

%%%%%%%%%%%%%%%%%%%%%%%%%%%%%%%%%%%%%%%%%%%%%%%%%%%%%%%%%%%
%%%%%%%%%%%%%%%%%%%%%%%%%%%%%%%%%%%%%%%%%%%%%%%%%%%%%%%%%%%
\section{Remaining Proofs from Section \ref{sec:fw:relations}}
%%%%%%%%%%%%%%%%%%%%%%%%%%%%%%%%%%%%%%%%%%%%%%%%%%%%%%%%%%%
%%%%%%%%%%%%%%%%%%%%%%%%%%%%%%%%%%%%%%%%%%%%%%%%%%%%%%%%%%%

%\ifLongVersion\proof{
\bigskip\noindent{\bf Proof of Lemma \ref{biquadratic:schemes}:}
Lemma 3 in \cite{arxiv-vmcai14} proves a general result for minimal-weight paths in weighted digraphs and the definition of a biquadratic path scheme hence refers to the cardinality of the set of vertices instead of the number of variables $N$. A mapping from our notions of (minimal) paths in $\mathcal{G}_R^n$ and path schemes to the general setting in \cite{arxiv-vmcai14} is via a technique called {\em zigzag automata} (weighted finite automata which can be viewed as digraphs). We refer an interested reader to \cite{arxiv-vmcai14} for the definition of zigzag automata. We only make a remark that by construction, zigzag automata of one-directional difference bounds relations have the number of control states bounded by $N$, from which our result follows.
\qed
%}\fi

%\ifLongVersion\proof{
\bigskip\noindent{\bf Proof of Lemma \ref{scheme:representants:ptime}:}
Consider the graph $\mathcal{G}_R^{N^4}$. Clearly, $\mathcal{G}_R^{N^4}$ has $N^5$ vertices, contains only forward edges and therefore no cycles. We will next compute $\minpathOp(i,j,n)$, a minimal weight path from $\xxki{0}{i}$ to $\xxki{n}{j}$ in $\mathcal{G}_R^{N^4}$, for each $1\leq n\leq N^4$ and $1\leq i,j\leq N$. This computation can be done iteratively, first for $n=0$ and $n=1$ (note that $\epsilon$ denotes the empty path):
\[
 \minpathOp(i,j,0) \stackrel{def}{=} \epsilon 
\]
\[
 \minpathOp(i,j,1)
 \stackrel{def}{=} \left\{\begin{array}{ll}
  \emptyset & \textrm{if } (\xxki{0}{i}\arrow{c_{ij}}{}\xxki{1}{j})\not\in\mathcal{G}_R^{N^4} \\
  \xxki{0}{i}\arrow{c_{ij}}{}\xxki{1}{j} & \textrm{otherwise}
 \end{array}\right.
\]
and then for each $n=2,\dots,N^4$ as:
\[
 \minpathOp(i,j,n)
 \stackrel{def}{=} \min \left\{
  \pi.\pi' ~|~ \exists 1\leq k\leq N ~.~ \bigwedge\begin{array}{l}
   \pi=\minpathOp(i,k,n-1) \\
   \pi'=\minpathOp(k,j,1)
 \end{array}\right\}
\]
where the $\min$ operator is defined as
\[
 \min(S) \stackrel{def}{=} \left\{\begin{array}{ll}
  \emptyset & \textrm{if } S=\emptyset \\
  \textrm{any }\rho\in S \textrm{ s.t. } \forall \rho'\in S ~.~ \weight{\rho}\leq\weight{\rho'} & \textrm{otherwise}
 \end{array}\right.
\]
Correctness of this computation can be shown by induction on $n$. The running time of each iteration is of the order $\mathcal{O}(N^3)$ and hence the total running time is of the order $\mathcal{O}(N^7)$. Then, we have:
\[ 
 \Pi_{ijkpq}\neq\emptyset \iff 
 \bigwedge\left\{\begin{array}{l}
  \minpathOp(k,k,p)\neq\emptyset \\
  \exists 0\leq r\leq q ~.~ \minpathOp(i,k,r)\neq\emptyset \wedge \minpathOp(k,j,q-r)\neq\emptyset
 \end{array}\right.
\]
If $\Pi_{ijkpq}\neq\emptyset$, we can define $\theta_{ijkpq} \stackrel{def}{=} \sigma.\lambda^*.\sigma'$ where $\lambda=\minpathOp(k,k,p)$ and
\[
 \sigma.\sigma' \stackrel{def}{=} \min\{ \nu.\nu' ~|~ \exists 0\leq r\leq q ~.~ \nu=\minpathOp(i,k,r), \nu'=\minpathOp(k,j,q-r) \}
\]
Since $0\leq r\leq N^4$, $\sigma.\sigma'$ can be computed in $\mathcal{O}(N^4)$ time. Since the ranges of $i,j,k,p,q$ are $1\leq i,j,k\leq N$, $0\leq p\leq N$, $0\leq q\leq N^4$, the computation of $\{ \theta_{ijkpq} ~|~ \Pi_{ijkpq}\neq\emptyset \}$ is of the order $\mathcal{O}(N^{12})$.
\qed
%}\fi

%\ifLongVersion\proof{
\bigskip\noindent{\bf Proof of Theorem \ref{th:cf:forward}:}
The correctness argument follows directly from Lemma \ref{computing:powers}, Lemma \ref{biquadratic:schemes}, from the definitions of $\theta_{ijkpq}$, $S_{ij}$, $\phi_{ij}$, and from \eqref{eq:cf:forward}. The complexity argument follows directly from Lemma \ref{scheme:representants:ptime} and definitions of $S_{ij}$, $\phi_{ij}$, and from \eqref{eq:cf:forward}.
\qed
%}\fi

%%%%%%%%%%%%%%%%%%%%%%%%%%%%%%%%%%%%%%%%%%%%%%%%%%%%%%%%%%%
%%%%%%%%%%%%%%%%%%%%%%%%%%%%%%%%%%%%%%%%%%%%%%%%%%%%%%%%%%%
\section{Remaining Proofs from Section \ref{sec:normalization}}
%%%%%%%%%%%%%%%%%%%%%%%%%%%%%%%%%%%%%%%%%%%%%%%%%%%%%%%%%%%
%%%%%%%%%%%%%%%%%%%%%%%%%%%%%%%%%%%%%%%%%%%%%%%%%%%%%%%%%%%

%\ifLongVersion\proof{
\bigskip\noindent{\bf Proof of Proposition \ref{normalized:characterization}:}
($``\Rightarrow``$) If the condition of Def. \ref{def:normalized} holds, it follows, since every segment is subpath of $\rho$ on positions $\{N^2,\dots,n-N^2\}$, that $\lcorners{\xi}=\emptyset$ for each $\xi\in\segments{\rho}$.
($``\Leftarrow``$) To see that the converse holds, notice that every subpath $\theta$ of $\rho$ at positions $\{N^2,\dots,n-N^2\}$ that is a long corner is also a subpath of some segment $\xi\in\segments{\rho}$. Then clearly, $\lcorners{\xi}\neq\emptyset$.
\qed
%}\fi

%\ifLongVersion\proof{
\bigskip\noindent{\bf Proof of Proposition \ref{unique:first:corner}:}
Assume that $\rho$ is of the form $\rho=\xxki{k_0}{i_0}\arrow{}{}\dots\arrow{}{}\xxki{k_m}{i_m}$. Consider the following set of indices:
\[ I \stackrel{def}{=} \{ q ~|~ (\xxki{k_p}{i_p}\arrow{}{}\dots\arrow{}{}\xxki{k_q}{i_q})\in\lbcorners{\rho} \} \]
Since $\lbcorners{\rho}\neq\emptyset$, it follows that $I\neq\emptyset$ and hence we can define $d \stackrel{def}{=} \min(I)$. By the definition of $I$ there exists a corner $\theta\in\lbcorners{\rho}$ of the form
\[ \theta = \xxki{k_a}{i_a}\arrow{}{}\dots\arrow{}{}\xxki{k_d}{i_d} \]
for some $0\leq a<d$. We next show that this corner is unique. By contradiction, suppose that there exists a corner $\theta'\in\lbcorners{\rho}$ of the form
\[ \theta' = \xxki{k_b}{i_b}\arrow{}{}\dots\arrow{}{}\xxki{k_d}{i_d} \]
for some $0\leq b<d$ such that $a\neq b$. Clearly, $k_a=k_b=k_d$. Suppose that $a<b$ (the case $b<a$ is symmetric), then $\theta'$ is a subpath of $\theta$ and hence $k_a\in\{k_{a+1},\dots,k_b=k_a,\dots,k_{d-1}\}$. Thus, $\theta$ is not basic, by the definition of a basic corner, contradiction.

Finally, we prove that $\extent{\theta}=N^2+1$. By contradiction, suppose that $e$ is the extent of $\theta$ such that $e>N^2+1$. By Proposition \ref{basic:subcorner}, there exists a basic corner $\theta'\in\bcorners{\theta}$ such that $\theta=\eta.\theta'.\eta'$ for some paths $\eta,\eta'$ and moreover, $\theta$ is of extent $N^2+1$ and starts at position $k'=k_0+e-(N^2+1)$. Since $e>N^2+1$, we have $k'>k_0$. Thus, $\eta$ and $\eta'$ are non-empty. Consequently, $\theta'$ is of the form $\xxki{k_b}{i_b}\arrow{}{}\dots\arrow{}{}\xxki{k_c}{i_c}$ for some $a<b<c<d$. Clearly, $c\in I$ and since $c<d$, it follows that $d$ is not minimal, contradiction.
% and 
%\[ \min(\positions{\theta'}) = \min(\positions{\theta}) + 1 = k_a + 1 \]
%Hence, $\theta'$ is of the form $\theta'=\xxki{k_a+1}{j_1} \arrow{}{}\dots\arrow{}{} \xxki{k_a+1}{j_2}$ for some $1\leq j_1,j_2\leq N$. Since $\theta'$ is a subpath of $\theta$, the above implies that $\theta=\eta.\theta'.\eta'$ for some non-empty paths $\eta,\eta'$. 
%Since $\extent{\theta'}=N^2+1$, we infer that $\theta'\in\lbcorners{\theta}\subseteq\lbcorners{\rho}$.
%Since  $\theta=\eta.\theta'.\eta'$, we obtain the contradiction on the minimality of $c$ and the choice of $\theta$, since if $d$ was indeed minimal, $\xxki{k_d}{i_d}$ would be the last vertex of $\theta'$ and not of $\theta$.
\qed
%}\fi

The following proposition proves that the shape of a segment $\xi$ determines whether the first lb-corner is right or left.
\begin{proposition} \label{corner:direction}
Let $n\geq 1$ and $0\leq p\leq q\leq n$ be integers, let $\rho$ be an extremal path in $\mathcal{G}_R^n$, and let $\xi\in\segments{\rho,p,q}$ be a segment such that $\lbcorners{\xi}\neq\emptyset$. 
Then, the first lb-corner of $\xi$ is right (left) if and only if $\xi$ starts at position $p$ ($q$, respectively).
\end{proposition}
\proof{
We assume that $\xi$ starts at position $p$ (a proof for the starting position $q$ is symmetric). 
%Clearly, each segment $\xi$ is of the form $\xi = \xxki{k}{i} \arrow{}{}\dots\arrow{}{} \xxki{\ell}{j}$ such that $k,\ell\in\{m,n-m\}$ and moreover, we have $\positions{\xi}\subseteq\{m,m+1,\dots,n-m\}$.
%
Let $\theta$ be the first lb-corner of $\xi$ and by contradiction, suppose that $\theta$ is left. By Proposition \ref{unique:first:corner}, $\theta$ has extent $N^2+1$. Thus, $\xi$ is of the form \[ \xi = \xxki{p}{i} \arrow{}{}\dots\arrow{}{} \underbrace{\xxki{\ell+N^2+1}{i_1} \arrow{}{}\dots\arrow{}{} \xxki{\ell}{i_2} \arrow{}{}\dots\arrow{}{} \xxki{\ell+N^2+1}{i_3}}_{\theta} \arrow{}{}\dots\arrow{}{} \xxki{k}{j} \] for some $1\leq i_1,i_2,i_3\leq N$ and some integers $k,\ell$. Without loss of generality, we assume that $\xxki{\ell}{i_2}$ is the first vertex of $\theta$ at position $\ell$. Since $\positions{\xi}\subseteq\{p,p+1,\dots,q\}$, it follows that $\ell\geq p$. Consequently, the subpath $\xxki{p}{i} \arrow{}{}\dots\arrow{}{} \xxki{\ell+N^2+1}{i_1}$ must traverse the position $\ell$. Let $\xxki{\ell}{i_0}$ be the last vertex of this subpath that is at position $\ell$. Then, $\xi$ can be written as \[ \xi = \xxki{p}{i} \arrow{}{}\dots\arrow{}{} \underbrace{\xxki{\ell}{i_0} \arrow{}{}\dots\arrow{}{} \xxki{\ell+N^2+1}{i_1} \arrow{}{}\dots\arrow{}{} \xxki{\ell}{i_2}}_{\theta'} \arrow{}{}\dots\arrow{}{} \xxki{\ell+N^2+1}{i_3} \arrow{}{}\dots\arrow{}{} \xxki{k}{j} \] Note that $\theta'$ is a lb-corner which contradicts the fact that $\theta'$ is the first lb-corner.
\qed}
%
%By Proposition \ref{corner:direction}, $\theta$ is a right corner.
%
\begin{proposition} \label{basic:corner:shape}
\textbf{(Shape of basic corners)}
Let $\rho$ be a basic right (left) corner such that $\extent{\rho}\geq 1$. Then, $\rho$ is of the form $\rho=\pi.\rho'.\pi'$ where $\pi$ is a forward (backward) edge, $\pi'$ is a backward (forward) edge, and $\rho'$ is a right (left) corner such that $\extent{\rho'}=\extent{\rho}-1$.
\end{proposition}
\proof{
We assume that $\rho$ is a basic right corner (a proof for a basic right corner is symmetric). Since $\extent{\rho}\geq 1$, $\rho$ consists of at least two edges and hence is of the form $\rho=\pi.\rho'.\pi'$ for some edges $\pi,\pi'$ and a path $\rho'$. 
Let $k$ be the position where $\rho$ starts and ends. Since $\rho$ is a right corner, then $\positions{\rho}=\{k,\dots,k+\extent{\rho}\}$. Since $\rho$ is basic, it follows that $k\not\in\positions{\rho'}$. Hence, $\pi$ ($\pi'$) ends (starts) at position $k+1$ and $\positions{\rho'}=\{k+1,\dots,k+\extent{\rho}\}$. Consequently, $\pi$ ($\pi'$) is forward (backward), $\rho'$ is a right corner that starts at position $k+1$, and $\extent{\rho'} = (k+\extent{\rho})-(k+1) = \extent{\rho}-1$.
\qed}
\begin{proposition} \label{basic:subcorner}
\textbf{(Basic subcorners)}
Let $\rho$ be a right corner of extent $e$ that starts at position $k$ and let $d\in\{1,\dots,e\}$ be an integer. Then, there exists a right corner $\rho'\in\bcorners{\rho}$ of extent $d$ such that $\rho'$ starts at position $k+e-d$.
% and moreover, $\rho=\eta.\rho'.\eta'$ for some paths $\eta,\eta'$.
%and \[ \min(\positions{\rho'}) = \min(\positions{\rho})+\extent{\rho}-d \]
\end{proposition}
\proof{
%Assume that $\rho$ is a right corner (the proof for left corners is symmetric). Suppose that $\rho$ starts and ends at position $k$ and let $e\stackrel{def}{=}\extent{\rho}$. 
Clearly, $\positions{\rho}=\{k,\dots,k+e\}$. Let $u$ and $w$ be the first and last vertex of $\rho$, respectively, and let $v$ be any vertex of $\rho$ at position $k+e$. Then, $\rho$ is of the form $u \arrow{}{}\dots\arrow{}{} v \arrow{}{}\dots\arrow{}{} w$. Consider the subpath $u \arrow{}{}\dots\arrow{}{} v$ and let $u'$ be the its last vertex at position $k+e-d$. Consider the subpath $v \arrow{}{}\dots\arrow{}{} w$ and let $w'$ be its first vertex at position $k+e-d$. We define $\rho' \stackrel{def}{=} u' \arrow{}{}\dots\arrow{}{} v \arrow{}{}\dots\arrow{}{} w'$. It is easy to see that $\rho'$ is a basic corner whose positions range between $k+e-d$ and $k+e$ and hence the extent of $\rho'$ is $d$.
\qed}

\begin{proposition} \label{lcorners:vs:lbcorners}
For each path $\rho$, $\lbcorners{\rho}=\emptyset$ if and only if $\lcorners{\rho}=\emptyset$.
\end{proposition}
\proof{
(``$\Leftarrow$'') Obvious, by the definition of $\lbcorners{\rho}$.
(``$\Rightarrow$'') We prove the converse. Suppose there exists $\theta\in\lcorners{\rho}$. Clearly, $\extent{\theta}>N^2$. By Proposition \ref{basic:subcorner}, there exists $\theta'\in\bcorners{\theta}\subseteq\bcorners{\rho}$ such that $\extent{\theta'}=\extent{\theta}>N^2$. Hence, $\theta'\in\bcorners{\rho}$ and $\theta'\in\lcorners{\rho}$. Consequently, $\theta'\in\lbcorners{\rho}$, by the definition of $\lbcorners{\rho}$.
\qed}

\begin{proposition} \label{pumping:rep:path:0}
Let $\mu$ be a repeating path such that $\card{\positions{\mu}}\leq N^2+1$. Then, $\lcorners{\mu^k}=\emptyset$ for all $k\geq 1$.
\end{proposition}
\proof{
We assume that $\mu$ is a forward path (the proof for $\mu$ being backward is symmetric). By contradiction, suppose that for some $k\geq 1$, $\mu^k$ contains a long right corner $\theta$ (the argument for a left corner is symmetric). Then, $\theta$ is of the form \[ \xxki{\ell}{j_0} \arrow{}{}\dots\arrow{}{} \xxki{\ell+N^2+1}{j_1} \arrow{}{}\dots\arrow{}{} \xxki{\ell}{j_2} \] for some $1\leq j_0,j_1,j_2\leq N$ and for some integer $\ell$. Clearly, there exist integers $1\leq p\leq q\leq r\leq k$ such that $\xxki{\ell}{j_0}$ ($\xxki{\ell+N^2+1}{j_1}$, $\xxki{\ell}{j_2}$, respectively) is traversed in the $p$-th ($q$-th, $r$-th, respectively) copy of $\mu$ in $\mu^k$. Since $\mu^k$ consists of $k$ copies of a forward repeating path $\mu$, it follows that the $r$-th copy traverses also $\xxki{\ell+N^2+1+(r-q)\cdot\llen{\mu}}{j_1}$, in addition to $\xxki{\ell}{j_2}$. Since $(r-q)\cdot\llen{\mu}\geq 0$, it follows that $\{\ell,\dots,\ell+N^2+1\} \subseteq \positions{\mu}$, which contradicts the assumption that $\card{\positions{\mu}}\leq N^2+1$. Hence, the claim holds.
\qed}

\begin{proposition} \label{fitting:subpath:in:corner}
Let $\rho$ be a right corner of the form $\rho=\xxki{k}{i}\arrow{}{}\dots\arrow{}{}\xxki{k}{j}$ and let $1\leq d\leq \extent{\rho}$. Then, $\rho$ has a subpath $\rho'$ of the form $\rho'=\xxki{k}{i}\arrow{}{}\dots\arrow{}{}\xxki{k+d}{p}$ for some $1\leq p\leq N$ such that $\positions{\rho'}=\{k,\dots,k+d\}$.
\end{proposition}
\proof{
We choose $\xxki{k+d}{p}$ as the first vertex of $\rho$ at position $k+d$ and define $\rho'$ as the subpath of $\rho$ that starts at $\xxki{k}{i}$ and ends at $\xxki{k+d}{p}$. Since $\rho$ is a corner and $\rho'$ its subpath, then $\positions{\rho'}\subseteq\positions{\rho}=\{k,\dots,k+\extent{\rho}\}$. By the choice of $\xxki{k+d}{p}$ and by the definition of $\rho'$, we have that $k+d\in\positions{\rho'}$ and $k+d+1\not\in\positions{\rho'}$. Consequently, $\positions{\rho'}=\{k,\dots,k+d\}$.
\qed}
%
%By Proposition \ref{basic:corner:shape}, we can decompose $\theta$ as $\theta=\tau.\theta'.\tau'$, where $\tau$ ($\tau'$) is a forward (backward) edge and $\theta'$ is a right corner with extent $N^2$.
%
\begin{proposition} \label{corner:decomposition}
\textbf{(Corner shortening / decomposition)}
Let $\rho$ be a right (left) corner such that $\extent{\rho}=N^2$. Then, there exists a compatible right (left) corner $\rho'$ such that either (i) $\extent{\rho'}<N^2$ or (ii) $\extent{\rho'}=N^2$ and $\rho'$ has subpaths $\eta,\mu,\tau,\mu',\eta'$ such that $\rho'=\eta.\mu.\tau.\mu'.\eta'$, $\mu$ is a forward (backward) repeating path, $\mu'$ is a backward (forward) repeating path, 
%$\eta$ is a forward (backward) path, $\eta'$ is a backward (forward) path, 
$\tau$ is a right (left) corner, $\llen{\eta}=\llen{\eta'}$, $1\leq \llen{\mu}=\llen{\mu'}\leq N^2$, and $\weight{\mu}+\weight{\mu'}<0$. Moreover, for all $k\geq 0$, $\eta.\mu^k.\tau.{\mu'}^k.\eta'$ is a right (left) corner and $\lbcorners{\eta.\mu^k}=\lbcorners{{\mu'}^k.\eta'}=\emptyset$.
\end{proposition}
\proof{
Suppose that $\rho$ is a right corner (a proof for left corner is symmetric). First, we assume that $\rho$ starts at position $0$. Then, since $\extent{\rho}=N^2$ there exist paths $\rho_1,\rho_2$ such that $\rho=\rho_1.\rho_2$, and $\rho_1$ ($\rho_2$) starts at position $0$ ($N^2$) and ends at position $N^2$ ($0$). Next, $\rho$ can be decomposed into the following form:
\[ \rho = (\underbrace{\xi_1.\xi_2\dots\xi_{N^2}}_{\rho_1}).(\underbrace{\zeta_{N^2}\dots\zeta_2\zeta_1}_{\rho_2}) \]
where for each $1\leq m\leq N^2$, $\xi_m$ and $\zeta_m$ are of the form:
\[ \xi_m = \xxki{m-1}{i_{m-1}} \arrow{}{}\dots\arrow{}{} \xxki{m}{i_m}   \hspace{10mm}  \zeta_m = \xxki{m}{j_m} \arrow{}{}\dots\arrow{}{} \xxki{m-1}{j_{m-1}} \]
(note that $\xxki{N^2}{i_{N^2}}=\xxki{N^2}{j_{N^2}}$) and moreover:
\begin{equation}\label{eq:corner:decomp:1} \positions{\xi_m},\positions{\zeta_m} \subseteq \{m-1,\dots,N^2\} \end{equation}
The property \eqref{eq:corner:decomp:1} can be achieved as follows. First, $\xxki{N^2}{i_{N^2}}$ can picked as the first vertex of $\rho$ at position $N^2$. This induces the above partition of $\rho$ into $\rho=\rho_1.\rho_2$. Then, for each $1\leq m<N^2$, the vertex $\xxki{m}{i_m}$ is chosen as the immediate successor of the last vertex of $\rho_1$ at position $m-1$. Symmetrically, for each $1\leq m<N^2$, the vertex $\xxki{m}{j_{m}}$ is chosen as the immediate predecessor of the first vertex of $\rho_2$ at position $m-1$.

Next, we define the sequence $\{(i_m,j_m)\}_{m=0}^{N^2}$. By the pigeonhole principle, there exist integers $0\leq a<b\leq N^2$ such that $(i_a,j_a)=(i_b,j_b)$. We define:
\[ \mu \stackrel{def}{=} \xi_{a+1}\xi_{a+2}\dots\xi_{b}  \hspace{10mm}  \mu' \stackrel{def}{=} \zeta_{b}.\zeta_{b-1}\dots\zeta_{a+1} \]
Note that $\mu$ ($\mu'$) is a forward (backward) repeating path of the form
\[ \mu = \xxki{a}{i_a} \arrow{}{}\dots\arrow{}{} \xxki{b}{i_a}  \hspace{10mm}  \mu' = \xxki{b}{j_a} \arrow{}{}\dots\arrow{}{} \xxki{a}{j_a} \]
and $\llen{\mu}=\llen{\mu'}=b-a\geq 1$. Clearly, there exist (possibly empty) paths $\eta,\tau,\eta'$ such that $\rho=\eta.\mu.\tau.\mu'.\eta'$. Note that $\eta,\tau,\eta'$ are of the form:
\[ \eta = \xi_{1}\xi_{2}\dots\xi_{a}  \hspace{10mm}  \tau = \xi_{b+1}\dots\xi_{N^2}.\zeta_{N^2}\dots\zeta_{b+1} \hspace{10mm}  \eta' = \zeta_{a}\zeta_{a-1}\dots\zeta_{1} \]
Note that $\tau$ traverses the vertex $\xxki{N^2}{i_{N^2}}$. Clearly, we have $\llen{\eta}=\llen{\eta'}$ and $1\leq \llen{\mu}=b-a=\llen{\mu'}\leq N^2$. It is easy to verify that \eqref{eq:corner:decomp:1} entails that 
$\tau$, $\mu^k.\tau.{\mu'}^k$, and $\eta.\mu^k.\tau.{\mu'}^k.\eta'$ are right corners, for all $k\geq 0$.
%we have $\positions{\eta.\tau.\eta'}\subseteq\{0,\dots,N^2\}$.

If $N^2\in\positions{\eta.\tau.\eta'}$ and $\weight{\mu}+\weight{\mu'}\geq 0$, we assign $\rho\leftarrow\eta.\tau.\eta'$ and apply the above construction again. We iterate this construction until either $N^2\not\in\positions{\eta.\tau.\eta'}$ or $\weight{\mu}+\weight{\mu'} < 0$. Note that this iteration terminates, since at each step, $\eta.\tau.\eta'$ traverses the position $N^2$ strictly less times than $\rho$ does; this is because $\llen{\mu}=\llen{\mu'}=b-a\geq 1$ and hence, all vertices in $\tau$ (including $\xxki{N^2}{i_{N^2}}$) are shifted by $b-a$ when constructing $\eta.\tau.\eta'$, hence the vertex $\xxki{N^2}{i_{N^2}}$ in $\tau$ will become $\xxki{N^2-(b-a)}{i_{N^2}}$ in $\eta.\tau.\eta'$.
Then, we define $\rho'$ to be the path from the last iteration. Either we have (i) $\extent{\rho'}<N^2$ (this happens when $N^2\not\in\positions{\rho'}$) or (ii) $\weight{\mu}+\weight{\mu'}<0$. We next prove that in case (ii), $\lbcorners{\eta.\mu^k}=\lbcorners{{\mu'}^k.\eta'}=\emptyset$, for all $k\geq 0$.

It follows from \eqref{eq:corner:decomp:1} and from the definition of $\mu$ that
\begin{equation}\label{eq:corner:decomposition:1} \positions{\eta.\mu^k.\tau.{\mu'}^k.\eta'} \subseteq \{ 0,\dots,N^2+\llen{\mu}\cdot k \} \end{equation}
for all $k\geq 0$ and hence $\eta.\mu^k.\tau.{\mu'}^k.\eta'$ is a right corner for all $k\geq 0$.
Next, we show that $\lbcorners{\eta.\mu^k}=\emptyset$ for all $k\geq 0$ (a proof of $\lbcorners{{\mu'}^k.\eta'}=\emptyset$ is symmetric). Since $\extent{\rho}=N^2$ and $\eta$, $\eta.\mu$ are subpaths of $\rho$, it follows that $\lbcorners{\rho}=\lbcorners{\eta.\mu}=\lbcorners{\eta}=\emptyset$. By \eqref{eq:corner:decomp:1} an by the definition of $\mu$, we have $\positions{\mu}\subseteq\{0,\dots,N^2\}$ and hence $\card{\positions{\mu}}\leq N^2+1$. Thus, by Proposition \ref{pumping:rep:path:0}, $\lbcorners{\mu^k}=\emptyset$. By contradiction, suppose that $\lbcorners{\eta.\mu^k}\neq\emptyset$ for some $k\geq 0$ and let $\theta$ be a corner from $\lbcorners{\eta.\mu^k}$. Since $\lbcorners{\eta}=\lbcorners{\mu^k}=\emptyset$, it follows that $\theta$ must traverse edges of both $\eta$ and $\mu^k$. Hence, $\in\positions{\theta}$. Since $\lbcorners{\eta.\mu}=\emptyset$, it follows that $k\geq 2$. By the definition of $\mu$, we have that $\mu^k$ starts at position $a$ and 
\begin{equation}\label{eq:corner:decomposition:2} \positions{\mu^k}\subseteq\{a,\dots,N^2+(b-a)\cdot\llen{\mu}\} \end{equation}
Consequently, $\theta$ ends at some position $c\geq a$. We prove that $\theta$ is a left corner. By contradiction, suppose that $\theta$ is a right corner. By the definition of right corners, it cannot be that $c>a$. By \eqref{eq:corner:decomposition:2}, we have $c\geq a$. Hence we infer that $c=a$. Thus, $\theta$ ends (and consequently also starts) at position $a$. Since $\theta$ contains at least one edge of both $\eta$ and $\mu^k$, the position $a$ is traversed at least three times, which contradicts the definition of a basic corner. Hence, $\theta$ is a left corner. Since $\positions{\eta}\subseteq\{0,\dots ,N^2\}$ and since $\theta$ contains at least one edge of $\eta$, it follows that $\theta$ starts at some position $c\in\{0,\dots,N^2\}$. Then, we infer, from the fact that $\theta$ is left and from \eqref{eq:corner:decomposition:1}, that $\positions{\theta}\subseteq\{0,\dots,c\}$. Since $c\leq N^2$, it follows that $\theta$ is short. Hence, $\lbcorners{\eta.\mu^k}=\lbcorners{{\mu'}^k.\eta'}=\emptyset$, for all $k\geq 0$.

Finally, note that the arguments above (for $0$ as the initial position of $\rho$) don't actually depend on the initial position of $\rho$ and hence generalize to corners that start at arbitrary position.
\qed}

%\ifLongVersion\proof{
\bigskip\noindent{\bf Proof of Lemma \ref{corner:decomposition:lemma:merged}:}
The result is obtained by chaining Proposition \ref{basic:corner:shape} and Lemma \ref{corner:decomposition:lemma:merged}. 
\qed
%}\fi

%%%%%%%%%%%%%%%%%%%%%%%%%%%%%%%%%%%%%%%%%%%%%%%%%%%%%%%%%%%
%%%%%%%%%%%%%%%%%%%%%%%%%%%%%%%%%%%%%%%%%%%%%%%%%%%%%%%%%%%
\section{Remaining Proofs from Section \ref{sec:closed:form:db}}
%%%%%%%%%%%%%%%%%%%%%%%%%%%%%%%%%%%%%%%%%%%%%%%%%%%%%%%%%%%
%%%%%%%%%%%%%%%%%%%%%%%%%%%%%%%%%%%%%%%%%%%%%%%%%%%%%%%%%%%

%\ifLongVersion\proof{
{\bf Proof of Proposition \ref{short:corner:replacement}:}
Let $n\geq 1$ and suppose that $\theta$ is a short right corner in $\mathcal{G}_R^n$, i.e. of the form $\theta=\xxki{k}{i}\arrow{}{}\dots\arrow{}{}\xxki{k}{j}$ such that $\positions{\theta}\subseteq\{k,\dots,k+N^2\}$. By Proposition \ref{balanced:shifting}, there is an isomorphic corner $\theta'=\xxki{0}{i}\arrow{}{}\dots\arrow{}{}\xxki{0}{j}$ such that $\positions{\theta'}\subseteq\{0,\dots,N^2\}$, i.e. $\theta'$ appears in $\mathcal{G}_R^n$. Hence, by Lemma \ref{computing:powers}, we have: 
\[ \exists \x' ~.~ R^{N^2}(\x,\x') \Rightarrow x_i-x_j\leq\weight{\theta} \] 
we infer (due to the conjunct $S_{fw}$ of $R_s$) that $R_s \Rightarrow x_i-x_j\leq\weight{\theta}$ and consequently, that $\mathcal{G}_{R_s}$ contains an edge $x_i\arrow{c}{}x_j$ for some $c\leq\weight{\theta}$. 

Next, we prove that $\theta'\stackrel{def}{=}\xxki{k}{i}\arrow{c}{}\xxki{k}{j}$ is an edge in $\mathcal{G}_{R_s}^n$. If $k=n$, $\theta$ is contains only vertical edges and consequently, $\mathcal{G}_R$ has an implied edge $x_i'\arrow{c}{}x_j'$, $c\leq\weight{\theta}$, and hence $\mathcal{G}_R^n$ (and consequently $\mathcal{G}_{R_s}^n$) has an edge $\xxki{n}{i}\arrow{c}{}\xxki{n}{j}$. If $k<n$, since $\mathcal{G}_{R_s}^n$ consists of $n$ identical copies of $\mathcal{G}_{R_s}$ and $\theta'=x_i\arrow{c}{}x_j$ is an edge in $S_{fw}$, it follows that $\mathcal{G}_{R_s}^n$ has an edge $\xxki{\ell}{i}\arrow{c}{}\xxki{\ell}{j}$ for each $\ell=0,\dots,k-1$.
Finally, observe that the conjunct $S_{bw}$ of $R_s$ allows us to make a symmetric argument about short left corners.
\qed
%}\fi

%\ifLongVersion\proof{
\bigskip\noindent{\bf Proof of Proposition \ref{encoding:of:segments}:}
Let $n\geq 1$ and $\xi$ be an extremal path in $\mathcal{G}_R^n$, i.e. a path of the form $\xxki{p}{i}\arrow{}{}\dots\arrow{}{}\xxki{q}{j}$ for some $p,q\in\{0,n\}$. We next analyze the four cases induced by values of $p,q$.
If $p=q=0$, then $\xi$ is a (short) corner and hence, by Proposition \ref{short:corner:replacement}:
\begin{equation} \label{eq:enc:segment:1} S_{fw} \Rightarrow x_i-x_j\leq\weight{\xi} \end{equation}
If $p=q=n$, we symmetrically obtain that $S_{bw} \Rightarrow x_i-x_j\leq\weight{\xi}$. 

Next, suppose that $p=0$ and $q=n$, i.e. that $\xi$ is forward.
It is not difficult to show that $\xi$ can be written as a sequence of (i) right corners and (ii) forward edges. For each right corner, there is a compatible vertical edge in $\mathcal{G}_{R_s}^n$, by Proposition \ref{short:corner:replacement}. Thus, we obtain path $\xi'$ in $\mathcal{G}_{R_s}^n$ that is compatible with $\xi$ and is a sequence of (i) vertical edges and (ii) forward edges.
%that originate from constraints in $S_{fw}$ 
%that originate from constraints in $R$
%
Next, it is easy to see that every subpath of $\xi'$ which follows the pattern $(vertical\mbox{-}edge)^*.forward\mbox{-}edge$ can be replaced by a single (implied) forward edge. Similarly, the suffix of $\xi'$ of the form $forward\mbox{-}edge.(vertical\mbox{-}edge)^*$ can be replaced by an implied forward edge. Thus, there exists a path $\xi''$ in $\mathcal{G}_{R_s}^n$ which is compatible with $\xi'$ (and hence with $\xi$) and which is a sequence of forward edges only. Then, it is not hard to show that $\xi''$ appears also in $\mathcal{G}_{R_{fw}}^n$. 
Since $\xi''$ as an extremal path in $\mathcal{G}_{R_{fw}}^n$, we infer, by Lemma \ref{computing:powers}, that
\[ R_{fw}^n(\x,\x') \Rightarrow x_i-x'_j\leq\weight{\xi''}\leq\weight{\xi} \]
Supposing that $\widehat{R}_{fw}(\ell,\x,\x')$ is a closed forms of $R_{fw}$, we obtain that 
\begin{equation} \label{eq:enc:segment:2} \widehat{R}_{fw}(\ell,\x,\x')[n/\ell] \Rightarrow x_i-x_j\leq\weight{\xi} \end{equation}
If $p=n$ and $q=0$, i.e. $\xi$ is backward, we can make a symmetric argument.
Then, it follows from the above reasoning (from Eq. \eqref{eq:enc:segment:1}, \eqref{eq:enc:segment:2}, and the two symmetric cases) that $\phi(\ell,\x,\x')$ encodes all extremal paths in $\mathcal{G}_R^n$ that have no long corners.
\qed
%}\fi

%\ifLongVersion\proof{
\bigskip\noindent{\bf Proof of Proposition \ref{encoding:of:extremal:paths}:}
Let $n\geq 1$ and $\rho$ be a extremal path in $\mathcal{G}_R^{2N^2+n}$. By Theorem \ref{normalization}, we can assume, without loss of generality, that $\rho$ is normalized. Recall from Definition \ref{def:segments} that $\rho$ can be written as 
\begin{equation} \label{eq:enc:norm:paths:1} \rho=\sigma_1.\xi_1\dots\sigma_m.\xi_m.\sigma_{m+1} \end{equation}
for some $m\geq 1$ where $\{\xi_1,\dots,\xi_m\}=\segments{\rho,N^2,N^2+n}$
and every segment $\xi_k$, $1\leq k\leq m$, is of the form:
\[ \xi_k = \xxki{N^2+p_k}{i_k}\arrow{}{}\dots\arrow{}{}\xxki{N^2+q_k}{j_k} \textrm{ where } 1\leq i_k,j_k\leq N, p_k,q_k\in\{0,n\} \]
Since $\positions{\xi_k}\subseteq\{N^2,\dots,N^2+n\}$, there exists, by Proposition \ref{balanced:shifting}, an isomorphic path $\xi_k'$ in $\mathcal{G}_R^{2N^2+n}$ of the form $\xi_k'=\xxki{p_k}{i_k}\arrow{}{}\dots\arrow{}{}\xxki{q_k}{j_k}$ such that $\positions{\xi_k'}\subseteq\{0,\dots,n\}$, i.e. $\xi_k'$ appears in $\mathcal{G}_R^n$. Since, $p_k,q_k\in\{0,n\}$, it follows that $\xi_k'$ is extremal in $\mathcal{G}_R^n$.
Since $\xi'$ contain no long corners, by definition \ref{def:normalized}, and is an extremal path in $\mathcal{G}_R^n$, we infer, by Proposition \ref{encoding:of:segments}, that $\xi'$ (and thus $\xi$) is encoded in $\phi(\ell,\y,\z)$.

It follows from Definition \ref{def:segments} that each $\sigma_k$, $1\leq k\leq m+1$,  appears in the first $N^2$ or in the last $N^2$ copies of $\mathcal{G}_R$, formally:
\[ \positions{\sigma_k}\subseteq\{0,\dots,N^2\} \textrm{ or } \positions{\sigma_k}\subseteq\{N^2+n,\dots,2N^2+n\} \]
It is easy to see that $\sigma_k$, $1\leq i\leq m+1$, is encoded either in $R^{N^2}(\x,\y)$ or in $R^{N^2}(\z,\x')$. 

E.g., suppose that $\rho=\sigma_1.\xi_1.\sigma_2.\xi_2.\sigma_3$ where
\[\begin{array}{lclclcl}
 \sigma_1 &=& \xxki{0}{i_1}\arrow{}{}\dots\arrow{}{}\xxki{N^2}{i_2}
  &\hspace{6mm}&
  \xi_1 &=& \xxki{N^2}{i_2}\arrow{}{}\dots\arrow{}{}\xxki{N^2+n}{i_3} \\
 \sigma_2 &=& \xxki{N^2+n}{i_3}\arrow{}{}\dots\arrow{}{}\xxki{N^2+n}{i_4}
  &\hspace{6mm}&
  \xi_2 &=& \xxki{N^2+n}{i_4}\arrow{}{}\dots\arrow{}{}\xxki{N^2}{i_5} \\
 \sigma_3 &=& \xxki{N^2}{i_5}\arrow{}{}\dots\arrow{}{}\xxki{0}{i_6}
\end{array}\]
Then, $\sigma_1,\xi_1,\sigma_2,\xi_2,\sigma_3$ are encoded as:
\[\begin{array}{lclclcl}
 R^{N^2}(\x,\y) &\Rightarrow& x_{i_1}-y_{i_2}\leq\weight{\sigma_1} 
  &\wedge&
  \phi(\ell,\y,\z)[n/\ell] &\Rightarrow& y_{i_2}-z_{i_3}\leq\weight{\xi_1} ~\wedge \\
 R^{N^2}(\z,\x') &\Rightarrow& z_{i_3}-z_{i_4}\leq\weight{\sigma_2} 
  &\wedge&
  \phi(\ell,\y,\z)[n/\ell] &\Rightarrow& z_{i_4}-y_{i_5}\leq\weight{\xi_2} ~\wedge \\
 R^{N^2}(\x,\y) &\Rightarrow& y_{i_5}-x_{i_6}\leq\weight{\sigma_3}
\end{array}\]
By transitivity, we infer that $\rho$ is encoded in $\psi$:
\[ \psi(\ell,\x,\x')[n/\ell] \Rightarrow x_{i_1}-x_{i_6}\leq \weight{\sigma_1}+\weight{\xi_1}+\weight{\sigma_2}+\weight{\xi_2}+\weight{\sigma_3}=\weight{\rho} \]

In general, for a path $\rho$ of the form \eqref{eq:enc:norm:paths:1}, we obtain $2m+1$ constraints implied by $R^{N^2}(\x,\y) \wedge \phi(\ell,\y,\z) \wedge R^{N^2}(\z,\x')$, one constraint for each path $\tau_1,\xi_1,\dots,\tau_m,\xi_m,\tau_{m+1}$. Then, by transitivity of these constraints, we infer that $\rho$ is encoded in $\psi$.
\qed
%}\fi

%\ifLongVersion\proof{
\bigskip\noindent{\bf Proof of Theorem \ref{balanced:closed:form}:}
We need to prove that for all $n\geq 1$:
\begin{equation} \label{eq:closed:form:prop}
 \widehat{R}(k,\x,\x')[n/k] \iff R^n(\x,\x')
\end{equation}
It is easy to verify that \eqref{eq:closed:form:prop} holds for each $n=1,\dots,2N^2$. Similarly, one can check that for every $n>2N^2$, \eqref{eq:closed:form:prop} is equivalent to 
\[ \psi(\ell,\x,\x')[n-2N^2/\ell] \iff R^n(\x,\x') \]
Equivalently, we can prove that for all $n\geq 1$:
\begin{equation} \label{eq:closed:form:prop:long}
 \psi(\ell,\x,\x')[n/\ell] \iff R^{2N^2+n}(\x,\x')
\end{equation}
By Proposition \ref{encoding:of:extremal:paths}, every extremal path in $\mathcal{G}_R^{2N^2+n}$ is encoded in $\psi(\ell,\x,\x')[n/k]$. Consequently, by Lemma \ref{computing:powers}, we have that $\psi(\ell,\x,\x')[n/k] \Rightarrow R^{2N^2+n}(\x,\x')$.
It remains to show that the $``\Leftarrow``$ implication of \eqref{eq:closed:form:prop:long} holds too. The intuition behind this is that for the strengthened relation $R_s$, one can prove that for all $n\geq 0$:
\begin{equation}\label{eq:strengthening:equivalence}
 R^{2N^2+n} \iff \underbrace{R^{N^2}}_{\textrm{prefix}} \circ R_s^n \circ \underbrace{R^{N^2}}_{\textrm{suffix}}
\end{equation}
Intuitively, this is because the strengthening of $R$ with $\exists \x' . R^{N^2}$ (with $\exists \x . R^{N^2}$) gives only constraints which are already implied by the above suffix (prefix) $R^{N^2}$. By the definition of $R_s$, we have that $R_s\Rightarrow S_{fw}(\x)$ and $R_s\Rightarrow S_{bw}(\x')$. It follows from the definition of $R_{fw}$ that $R_s^n \Rightarrow R_{fw}^n$. Similarly, $R_s^n \Rightarrow R_{bw}^n$. Thus, we have:
\begin{equation}\label{eq:strengthening:equivalence:2}
\begin{array}{lcl}
 R^{N^2} \circ R_s^n \circ R^{N^2} 
  &\Rightarrow& R^{N^2} \circ [R_{fw}^n \wedge R_{bw}^n \wedge S_{fw}(\x) \wedge S_{bw}(\x')] \circ R^{N^2} \\
  &\iff& \psi(\x,\x',\ell)[n/\ell]
\end{array}
\end{equation}
The last equivalence above is by the definition of $\psi$. Combining \eqref{eq:strengthening:equivalence} with \eqref{eq:strengthening:equivalence:2}, we infer that the $``\Leftarrow``$ implication of \eqref{eq:closed:form:prop:long} also holds.
\qed
%}\fi

%%%%%%%%%%%%%%%%%%%%%%%%%%%%%%%%%%%%%%%%%%%%%%%%%%%%%%%%%%%
%%%%%%%%%%%%%%%%%%%%%%%%%%%%%%%%%%%%%%%%%%%%%%%%%%%%%%%%%%%
\section{Remaining Proofs from Section \ref{sec:octagons}}
%%%%%%%%%%%%%%%%%%%%%%%%%%%%%%%%%%%%%%%%%%%%%%%%%%%%%%%%%%%
%%%%%%%%%%%%%%%%%%%%%%%%%%%%%%%%%%%%%%%%%%%%%%%%%%%%%%%%%%%

%\ifLongVersion\proof{
\bigskip\noindent{\bf Proof of Lemma \ref{oct:consistent:powers}:}
See Eq. (4.16) in the proof of Proposition 15 in \cite{arxiv-lmcs}.
\qed
%}\fi

%\ifLongVersion\proof{
\bigskip\noindent{\bf Proof of Lemma \ref{min:inconsistent}:}
If a relation $R$ is not $*$-consistent, there exists a computable exponential upper bound $B$ on $K_R$ of the order $2^{\mathcal{O}(\bin{R})}$ (see Lemma 20 in \cite{arxiv-vmcai14}). Since the $n$-th power of an octagonal relation can be computed in $\mathcal{O}(\bin{R}\cdot\log_2n)$ time (see Lemma 18 in \cite{arxiv-vmcai14}), one can find $K_R$ by performing a binary search on the interval $\{1,..,B\}$ and use the fact that the powers of $R$ can be computed by fast exponentiation. 
\qed
%}\fi

%\ifLongVersion\proof{
\bigskip\noindent{\bf Proof of Theorem \ref{closed:form:octagons}:}
First, suppose that $R$ is $*$-consistent. Then, for each $n\geq 1$, we have that $R^n(\x,\x')$ is consistent and hence:
\[\begin{array}{lclcl}
 \widehat{R}(k,\x,\x')[n/k] 
   &\iff& \widehat{\overline{R}}(k,\y,\y')[n/k][\sigma] &\hspace{6mm}& \textrm{(by \eqref{eq:cf:octagons})}
 \\
   &\iff& \overline{R}^n(\y,\y')[\sigma]
 \\
   &\iff& R^n(\x,\x') && \textrm{(by Lemma \ref{oct:consistent:powers})}
\end{array}\]
Next, suppose that $R$ is not $*$-consistent. For each $1\leq n<K_R$, we have that $R^n(\x,\x')$ is consistent and hence:
\[\begin{array}{lclcl}
 \widehat{R}(k,\x,\x')[n/k]
   &\iff& \widehat{\overline{R}}(k,\y,\y')[n/k][\sigma] \wedge n<K_R &\hspace{6mm}& \textrm{(by \eqref{eq:cf:octagons})}
 \\
   &\iff& \overline{R}^n(\y,\y')[\sigma] \wedge \true
 \\
   &\iff& R^n(\x,\x') && \textrm{(by Lemma \ref{oct:consistent:powers})}
\end{array}\]
For each $n\geq K_R$, we obtain:
\[\begin{array}{lclcl}
 \widehat{R}(k,\x,\x')[n/k]
   &\iff& \widehat{\overline{R}}(k,\y,\y')[n/k][\sigma] \wedge n<K_R &\hspace{6mm}& \textrm{(by \eqref{eq:cf:octagons})}
 \\
   &\iff& \false
\end{array}\]
Hence, in all cases, \eqref{eq:cf:octagons} defines a closed form of $R(\x,\x')$. The fact that $\widehat{R}(k,\x,\x')$ is PTIME-computable in the existential fragment of Presburger arithmetic follows immediately from Lemma \ref{min:inconsistent} and Corollary \ref{closed:form}.
\qed
%}\fi

%\section{todo list}
%\begin{itemize}
% \item lower bounds
% \item periodicity, prefix, period, examples with exponential prefix
% \item don't say "we present an algorithm"; instead, say "show that the closed form can be computed"
% \item more precise complexity results on relational composition: polynomial in N, n, ...
% \item todo: give details on complexity in fast exponentiation
%\end{itemize}

\end{document}